\documentclass[10pt,conference]{IEEEtran}
\IEEEoverridecommandlockouts

\usepackage{cite}
\usepackage{url}
\usepackage{xspace} 
\usepackage{amsmath,amssymb,amsfonts}
\usepackage{algorithm}
\usepackage{algpseudocode}
\usepackage{graphicx}
\usepackage{textcomp}
\usepackage{xcolor}
\usepackage{array} % table vertical central
\usepackage{makecell} % table make cell position
\usepackage{tabularx,booktabs}
\usepackage{multirow}

\begin{document}

\newcommand{\name}{\textsc{DISQ}\xspace}
\newcommand{\itname}{\textit{DISQ}\xspace}

\title{\name: \underline{D}ynamic \underline{I}teration \underline{S}kipping for Variational \underline{Q}uantum Algorithms}

\author{
Junyao Zhang\textsuperscript{1}, 
Hanrui Wang\textsuperscript{2}, 
Gokul Subramanian Ravi\textsuperscript{3,4},\\

Frederic T. Chong\textsuperscript{3},
Song Han\textsuperscript{2},
Frank Mueller\textsuperscript{5},
Yiran Chen\textsuperscript{1}\\
\textsuperscript{1}Duke University, \textsuperscript{2}MIT,
\textsuperscript{3}University of Chicago,
\textsuperscript{4}University of Michigan,
\textsuperscript{5}North Carolina State University\\
jz420@duke.com
}

\maketitle

\begin{abstract}
In the noisy intermediate scale quantum (NISQ) era, the Variational Quantum Algorithm (VQA) has emerged as one of the most promising approaches to harness the power of quantum computers. In VQA, a classical optimizer iteratively updates the parameters of a variational quantum circuit to minimize a cost objective obtained by executing the quantum circuit on real quantum hardware. However, the deployment of VQA applications on NISQ devices encounters substantial noise, which degrades training stability. Moreover, the drift of noise is particularly intractable due to its dynamic nature in duration and magnitude. Noise drift leads to significant deviations in VQA iteration's objective function estimation and shapes a dynamic noisy landscape, which poses a considerable challenge for stable VQA parameter training, thereby hampering the accurate convergence of VQA optimizations.

This paper proposes \name to craft a stable landscape for VQA training and tackle the noise drift challenge. \name adopts a ``drift detector" with a reference circuit to identify and skip iterations that are severely affected by noise drift errors. Specifically, the circuits from the previous training iteration are re-executed as a reference circuit in the current iteration to estimate noise drift impacts. 
The iteration is deemed compromised by noise drift errors and thus skipped if noise drift flips the direction of the ideal optimization gradient. 
To enhance noise drift detection reliability, we further propose to leverage multiple reference circuits from previous iterations to provide a well-founded judge of current noise drift. Nevertheless, multiple reference circuits also introduce considerable execution overhead. To mitigate extra overhead, we propose Pauli-term subsetting (prime and minor subsets) to execute only observable circuits with large coefficient magnitudes (prime subset) during drift detection. Only this minor subset is executed when the current iteration is drift-free.

Evaluations across various applications and QPUs demonstrate that \name can mitigate a significant portion of the noise drift impact on VQAs and achieve 1.51-2.24$\times$ fidelity improvement over the traditional baseline. 
\name's benefit is 1.1-1.9$\times$ over the best alternative approach while boosting average noise detection speed by 2.07$\times$.

\end{abstract}


\section{Introduction}
Quantum computing is a revolutionary computational model that is poised to leverage substantial quantum mechanical phenomena to provide computing advantages in resolving some classically intractable problems in domains, such as chemistry \cite{VQE_chemistry, chemistry}, biology \cite{biology}, fundamental software algorithms \cite{grover, shor}, and machine learning \cite{QML, qsvm}.

\begin{figure}[t]
  \centering
  \includegraphics[width=0.9\linewidth]{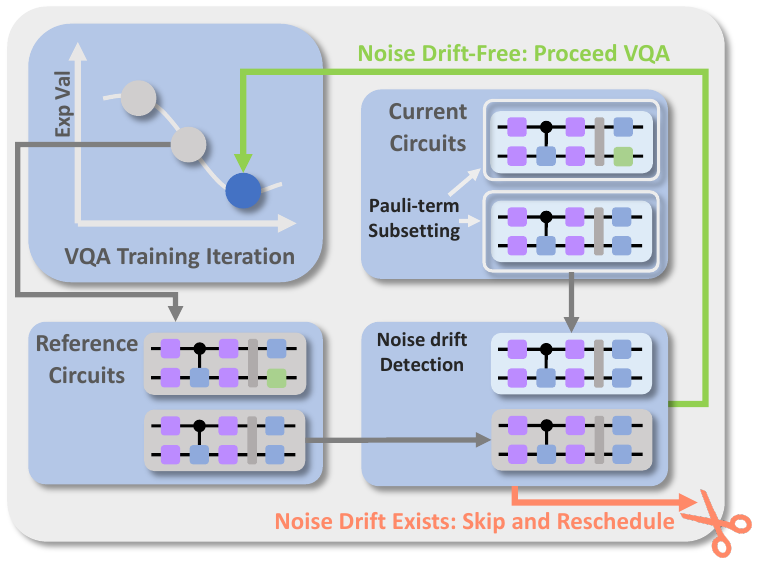}
  \caption{Computing resource-efficient iteration skipping approach to filter out the noise drift impact. Previous iterations act as optimal reference circuits to detect the noise drift on the current VQA iteration. Pauli-term subsetting is utilized to proactively minimize computation during noise drift detection.
  }
  \label{fig:teaser}
  \vspace{-15pt}
\end{figure}

One of the most promising noise intermediate scale quantum (NISQ) \cite{NISQ} algorithms that can provide a quantum advantage is the variational quantum algorithm (VQA) \cite{VQA, VQE}, which has been widely applied to critical applications in chemistry \cite{VQE_chemistry, VQE}, approximation \cite{QAOA}, and physics \cite{nucleus}, among others. VQA is a long-running iterative algorithm that deploys a classical optimizer to train a parameterized quantum circuit on a quantum machine. The quantum circuit parameters are tuned in each iteration to approach the application's targets, which are usually minimization problems, such as estimating the ground state energy of molecules.

In spite of quantum supremacy having been theorized in the aforementioned domains \cite{grover, shor, VQE, biology, QML, zhang2023c, zhang2023majority}, quantum processing units (QPU) in the contemporary NISQ era are still vulnerable to various types of noise, such as decoherence errors, gate errors, state preparation and measurement (SPAM) errors, and crosstalk. These noise errors stem from a multitude of sources, such as device defects \cite{decoherence_dielectric_loss}, thermal fluctuations \cite{fluctuations, decoherence}, magnetic flux \cite{fluctuations, Low_frequency, trapionchallenge, ZeemanIon}, qubit coupling destruction \cite{amorphous_solids}, insulation problems \cite{decoherence_dielectric_loss}, and other external stimuli \cite{external1, external2, externalerror}. Due to the dynamic nature of quantum systems and the current limitations of quantum devices fabrication \cite{crosstalk2}, these noise sources vary over time in intensity, thereby causing unpredictable fluctuations that deteriorate the quantum system.

With awareness of qubit restrictions in deploying full-scale quantum error correction (QEC) \cite{QEC, realization_QEC} and the perniciousness of noise, multiple works have investigated mitigation techniques for general quantum circuits \cite{miti_1, miti_2, miti_3, miti_4, miti_5, miti_7, miti_8, miti_9}. To harvest quantum computing power while maintaining robustness in the face of different forms of noise errors. \cite{VQA_miti, quantumnat, wang2022qoc} have proposed noise-aware training frameworks to boost VQA training speed and accuracy of results. \cite{quantumnas} have proposed noise-aware search for robust quantum ansatz. Although these works provide the potential to improve result fidelity or enhance training robustness, they primarily focus on static noise. Noise is assumed to be static or to remain stable for sufficient periods, such that these techniques can adequately capture and model the characteristics of static noise to provide mitigation methods. Unfortunately, focusing solely on static noise is insufficient for deploying VQA and harvesting its power on QPUs.

VQA iteratively tunes circuit parameters with the support of gradients \cite{optimizer}. The objective of the optimizer is to filter out the optimal gradient among disparate gradients to generate a corresponding set of parameters in each iteration, thereby achieving convergence towards the desired objective function. This convergence is expected to occur when the noise landscape of the quantum device is stable and consistent during the gradient estimation process, which allows the optimizer to accurately discern the most optimal circuit parameters. One of the fundamental challenges hampering the deployment of VQAs for practical utilization on QPUs is noise drift error originating from dynamic sources. The term ``noise drift'' refers to deviations in a quantum circuit's output distribution caused by shifts in the device characteristics of one or more qubits in the quantum circuit \cite{fluctuations, decoherence, decoherence_dielectric_loss, Low_frequency}. 
Noise drift, characterized by the unpredictable and time-varying nature of noise in quantum devices, poses a significant challenge to quantum algorithms, particularly for long-running iterative applications, such as VQA. Since noise drift alters the VQA tuning landscape over iterations, it leads to inconsistent gradient estimation each time and thus disrupts the convergence of VQA. This dynamic nature results in noise drift being hard to model and suppress with static noise mitigation techniques. 
The state-of-the-art method targeting dynamic noise, QISMET \cite{qismet}, predominantly addresses spike-like transient noise but performs inadequately in the presence of complex noise, such as noise drift, potentially misdirecting the VQA training.

In this paper, we propose \name, \underline{D}ynamic \underline{I}teration \underline{S}kipping for Variational \underline{Q}uantum Algorithms, a novel computing resource-efficient iteration skipping approach that crafts a reliable landscape for VQA applications by filtering out noise drift errors as in Fig.\ref{fig:teaser}. 
Drawing inspiration from QISMET in handling dynamic noise, (i) \name employs a previous iteration as an optimal reference to detect the noise drift for the current iteration by estimating the discrepancy between the reference output from the prior iteration and its result in the current iteration. Subsequently, \name adopts traditional per-iteration gradient calculation and combines it with noise drift error detection to estimate machine-obtained and drift-free gradients. A VQA iteration is accepted only if the direction of the machine-observed gradient loosely matches the direction of the drift-free gradient; 
Advancing beyond QISMET, (ii) \name augments multiple previous iterations as references to further enhance noise drift detection reliability in facing intractable noise drift. Although multi-reference circuits aid in shaping the VQA training landscape, they also introduce extra execution costs;
(iii) To minimize the extra execution cost, \name groups the Pauli terms of the target Hamiltonian by Pauli-term subsetting (prime and minor subsets) and partitions the execution. Prime subsets (Pauli terms with dominating coefficients) of references and prime subsets corresponding to the current iteration are executed during noise drift detection. The minor subset for the current iteration is executed only if noise drift is not detected. Subsetting conserves computation in noise drift detection, and execution partitioning eliminates unnecessary computation in skipping iterations. Evaluation demonstrates that \name achieves 1.1-1.9$\times$ fidelity improvements over the best alternative approach while boosting average noise detection speed by 2.07$\times$. This paper thus makes the following contributions:
 \begin{itemize}
  \item $\textbf{\textit{\itname Framework}}$: We propose \name, a computing resource-efficient method to actively discover noise drift instances that severely impact VQA accurate convergence. \name controls VQA iterations to actively skip noise drift errors and maintain VQA optimization under reliable scenarios.

  \item $\textbf{\textit{Noise Drift Detection with Multi-References}}$: We leverage multiple previous iterations as references to faithfully detect noise drift errors in VQA iterations, enabling precise estimation of noise drift impacts with \name.
  
  \item $\textbf{\textit{Concept and Design of Pauli-term Subsetting}}$: We introduce a novel perspective on quantum circuits for noise detection: instead of employing all the observable circuits in a brute-force manner, Pauli-term subsetting groups the dominant observable circuits (prime subset) as effective substitutes for noise detection. Subsetting proactively enables execution acceleration in \name even with numerous references.

  \item $\textbf{\textit{Robust Landscape with Reasonable Overhead}}$: Through the aforementioned steps, \name effectively filters a substantial portion of noise drift's impact on VQAs, attaining 1.51-2.24$\times$ fidelity increase over traditional baseline. \name surpasses the best alternative approach with a 1.1-1.9$\times$ benefit while enhancing average noise detection speed by 2.07$\times$.
  
\end{itemize}

\section{BACKGROUND}
\subsection{Noise in NISQ QPUs} \label{sec:noise_drift}
In the era of NISQ, two primary technologies in quantum architectures: superconducting transmon qubits \cite{superconducting} and trapped-ions \cite{trapionchallenge, ionq} are being pursued for universal quantum computing. Despite the fact that no definitive verdict has been reached in their performance comparison \cite{quantumArchcomp}, both require precise control due to their sensitivity to various types of noise. The scalability of QPUs built using these two technologies is limited by noise errors, such as 
(i) Decoherence error: natural decay caused by the energy exchange between a qubit and its environment, which makes the qubit lose its quantum properties (state) over time; 
(ii) State preparation and measurement (SPAM) errors: caused by imprecise state initialization and measurement \cite{SPAM};
(iii) Gate error: imperfect gate operations, such as depolarization, which are approximately 0.1\% and 4\% for 1-qubit gates and 2-qubit gates, respectively, for IBMQ \cite{gate_error};
(iv) Device-specific error: QPU-specific noise profiles that vary in spatial and temporal noise, arising from factors like topology cross-talk \cite{crosstalk1, crosstalk2}, inhomogeneous qubits during manufacture \cite{IontrapExp1, IontrapExp2, amorphous_solids}, imperfections in gate implementation and control, or specific external interference \cite{external1, external2}, where the latter is a significant source of error such as magnetic fields affecting Zeeman trapped-ion qubits \cite{ZeemanIon, trapionchallenge} and unstable near-resonant two-level systems affecting transmon qubits \cite{decoherence}. 

\begin{figure}[t]
  \centering
  \includegraphics[width=\linewidth]{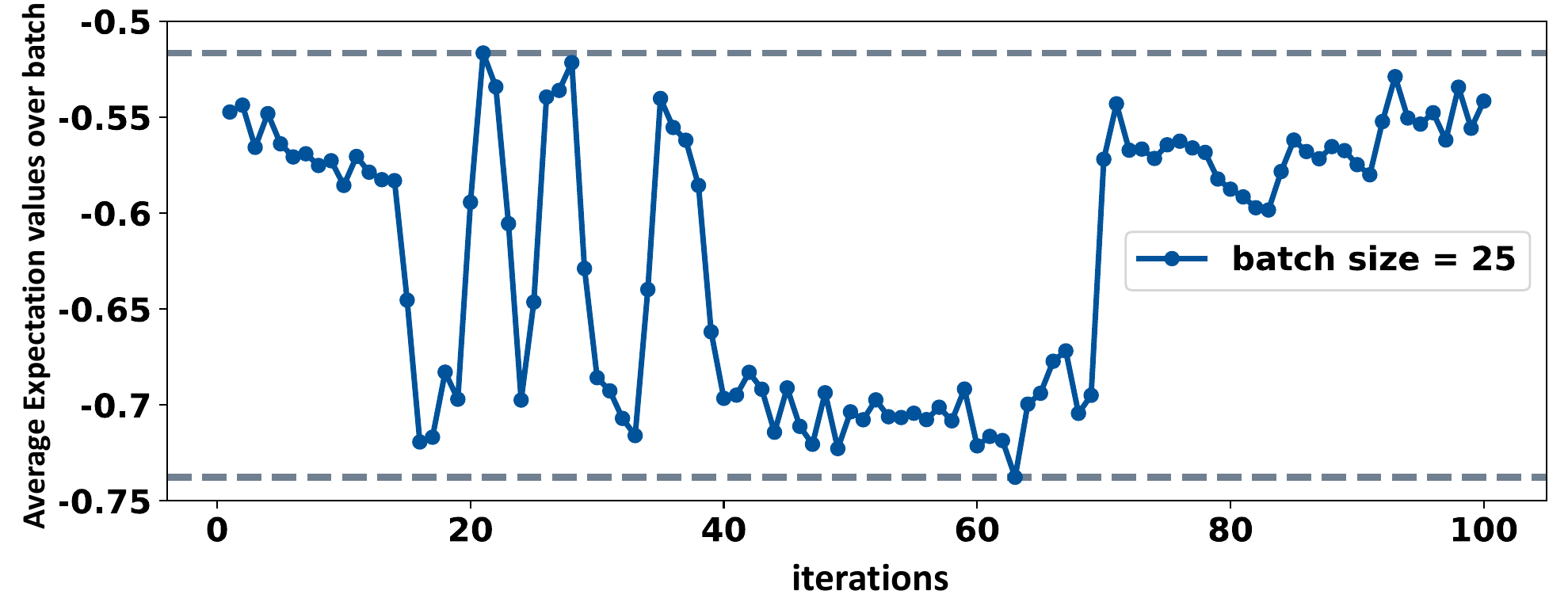}
  \vspace{-20pt}
  \caption{Noise drift errors on circuits. Circuit data are collected by 100 continuous runs of a circuit batch from an experiment on IBMQ Belem. Each data point is the average expectation value of the circuit batch (25 identical circuits). The mean value is -0.62, but the range value is concerning 0.22.}
  \label{fig:noise_drift}
  \vspace{-1.5em}
\end{figure}

The presence of those stochastic errors and environmental variations renders the noise drift to be dynamic in nature \cite{amorphous_solids, fluctuations, Low_frequency, quantum_drift}. Further, the impacts of different noise drift errors compound with each other and accumulate over time, thereby increasing the overall probability of obtaining erroneous outcomes, particularly in long-running applications such as VQAs. Gradient evaluation discrepancies can mislead the tuning process in unfavorable directions, thereby impeding the convergence and accuracy of VQA results. Fig.\ref{fig:noise_drift} demonstrates a severe case of noise drift affecting circuit fidelity, obtained from IBMQ Belem. Each data point represents the average expectation value obtained over a circuit batch (25 identical circuits). Although the mean of expectation values over 100 executions of this batch is -0.62, its range is large at 0.22. This proves that the statistical robustness offered by executing multiple circuit shots (for capturing probabilistic output distributions) cannot compensate for noise drift errors. While such severe noise drift might not always occur, milder noise drifts that continuously steer circuit results away from the correct objective ground state are frequently encountered and still perturb the accuracy of results.

% variance: 0.004744994457211989
% std: 0.068883920164375
% range: 0.22115420567668997
% range: -0.622071376939934
% -0.5165, min:-0.7377, diff:0.2212, diff per: 0.4281

%%%%%%%%%%%%
\subsection{Variational Quantum Algorithms}\label{sec:VQA}

VQA, a hybrid quantum-classical algorithm, is widely used in chemistry and approximations such as Variational Quantum Eigensolver (VQE) \cite{VQE} and Quantum Approximate Optimization Algorithm (QAOA) \cite{QAOA}. The VQA problem can be generalized to a Hamiltonian, $\hat{\text{H}}$, a linear combination of Pauli terms $\hat{P_i}$, and their numerical coefficients $c_i$, to describe the total energy of the system, as shown in Equation \ref{eq:H}. 
\begin{equation} \label{eq:H}
\footnotesize
    \hat{\text{H}} = \sum_i c_i\hat{P_i}
\end{equation}
The objective function is to find the ground state energy of the system, which corresponds to the lowest eigenvalue, $\lambda_{min}$, of $\hat{\text{H}}$ \cite{VQE}.
To approximate the expectation value for $\hat{P_i}$, the parameterized quantum circuit (ansatz) with parameters $\overrightarrow{\theta}, \theta \subseteq \mathbb{R}$, are iteratively executed on QPUs and tuned by a classical optimizer. The expectation value is derived from ansatz measurements over different observables (bases). The objective function is represented by
\begin{equation} \label{eq:objective}
\footnotesize
% \begin{split}
      \text{min}\,\lambda({\overrightarrow{\theta}}) = \text{min} \, \langle\psi({\overrightarrow{\theta}}) |\hat{\text{H}}| \psi({\overrightarrow{\theta}})\rangle = 
  \text{min} \, \sum_ic_i\langle\psi({\overrightarrow{\theta}}) |\hat{P_i}| \psi({\overrightarrow{\theta}})\rangle
% \end{split}
\end{equation}
$|\psi({\overrightarrow{\theta}})\rangle$ is the eigenvector corresponding to $\lambda({\overrightarrow{\theta}})$. Ideally, the application monotonically converges to the minimum. However, machine noise, such as various noise drifts, can taint the gradient computation and thus bias the VQA optimization process. Therefore, VQA relies on  noise-robust optimizers to become noise resilience \cite{VQA, VQA_optimizer}.

\section{MOTIVATION}

\begin{figure}[b]
  \centering
  \vspace{-1.5em}
  \includegraphics[width=0.9\linewidth]{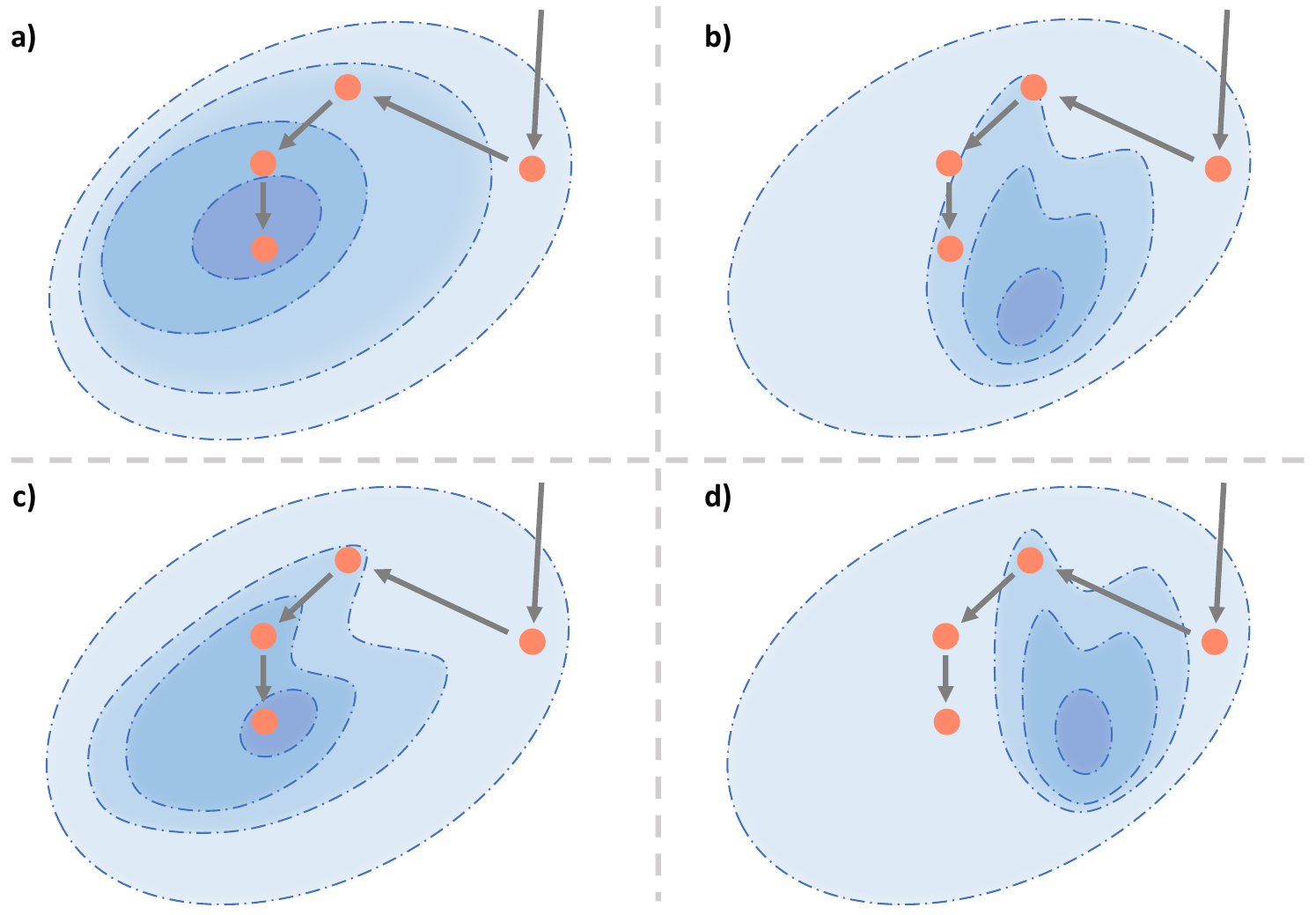}
  \caption{VQA landscape navigation. Contour levels correspond to different objective function values, with darker colors indicating smaller values. The orange dot marks the objective function value for each parameter configuration chosen by the optimizer. The tuning process is represented by an arrow trace. a) represents the drift-free (ideal) scenario; b) and d) depict scenarios with out-of-range noise drift; c) illustrates a scenario where noise drift is acceptable.
  }
  \label{fig:navi}
  % \vspace{-1.5em}
\end{figure}

\subsection{Dynamic Noise Landscape Navigation} \label{landscape}
Classical gradient-based optimizers in VQA are implemented in multiple ways to calculate the gradient based on the prior gradients and then tune the parameters of the ansatz to drive the objective function into a ground state. The underlying assumption of the optimizer is the evaluation of the gradient under the same noise environment.
\begin{figure*}[t]
  \centering
  \includegraphics[width=0.9\linewidth]{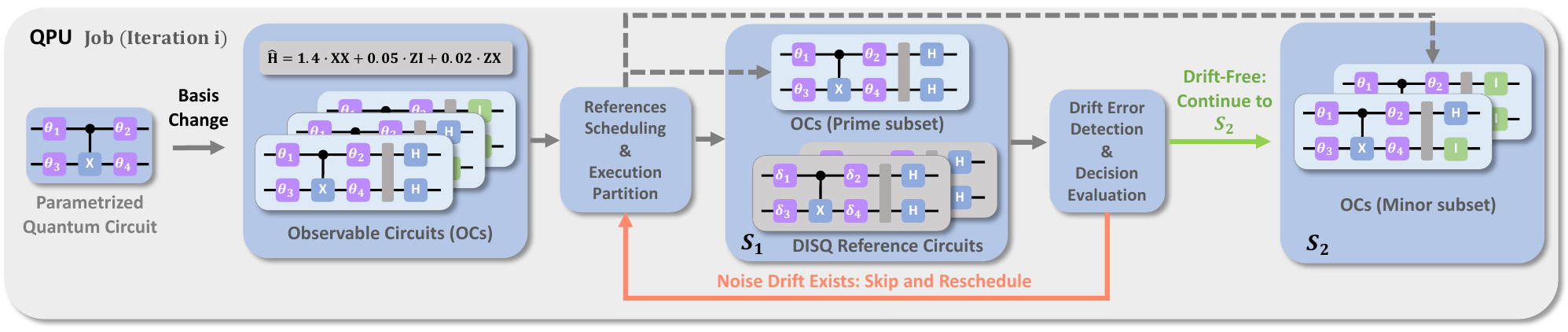}
  % {Figures/overview_new.pdf}
  \vspace{-10pt}
  \caption{Overview of \name: a VQA iteration execution is partitioned into two stages ($S_1$, $S_2$); Prime subset corresponding to the current iteration $i$ (blue circuit in $S_1$) and its references (gray circuits) are executed in $S_1$ to detect the noise drift; If noise drift is present, \name skips the results of the current job and reschedules all the circuits in $S_1$ via the next job (orange line). Otherwise, minor subset corresponding to the current iteration $i$ (blue circuits in $S_2$) are executed to proceed with VQA (green line).}
  \label{fig:overview}
  \vspace{-15pt}
\end{figure*}

To better understand the impact of noise drift in VQA, Fig.\ref{fig:navi} visualizes the VQA landscape for a minimization problem under different scenarios. Each contour level represents a value of the objective function, with a darker blue color indicating a smaller value in the objective function estimation. The orange dot represents the objective function value for each parameter configuration selected by the optimizer. Ideally, the VQA optimizer steadily tunes the parameters towards a minimum value of the objective function, which is represented by moving the orange point from the contour level in light blue to the deeper blue contour level, as shown in Fig.\ref{fig:navi}-a. However, in reality, the presence of noise drift alters contour levels' shape and position, leading to discrepancies between the machine-obtained and ideal values. Such discrepancies even vary with each iteration of the optimization process. In comparison with Fig.\ref{fig:navi}-a (drift-free), Fig.\ref{fig:navi}-b illustrates how optimizer tuning is disturbed by severe noise drift. Optimizers are incapable of finding the minimum value in such a landscape with substantial fluctuations in objective function estimation. Therefore, it becomes crucial to estimate noise drift errors and craft a stable environment for the gradient calculation of the optimizer. Specifically, optimizers only process VQA iterations under scenarios such as Fig.\ref{fig:navi}-c, where the machine-obtained contour levels are loosely aligned with the ideal (drift-free) contour levels in Fig.\ref{fig:navi}-a.

\subsection{Limitations in Dynamic Noise Estimation} \label{sec:detection}
Traditional VQA optimizers calculate the tuning gradient $G_{i}$ for iteration $(i)$ from the discrepancy between the machine-obtained objective function estimations (or energy) of iterations $(i)$ and $(i-1)$ to select the next iteration $(i+1)$ parameter: 
\begin{equation}\label{eq:G}
\footnotesize
\vspace{-0.5em}
  G_{i} = E_{i} - E_{i-1}
\end{equation}
where $E_i$ denotes the energy obtained by the quantum machine corresponding to iteration $(i)$. 

Regarding the cutting-edge transient mitigation technique QISMET \cite{qismet}, it selects the adjacent previous iteration $(i-1)$ as the reference circuit and re-executes it in the current iteration $(i)$ to estimate noise errors for iteration $(i)$. Noise impact is estimated by comparing the energy of the reference circuit repetition in the current iteration $Er_{i-1}$ with its previous execution $E_{i-1}$:
\begin{equation}\label{eq:D}
\footnotesize
    N_i = Er_{i-1} - E_{i-1}
\end{equation}
The noise-free energy $Ef_i$ is predicted by removing the noise error component from the energy estimation using Equation \ref{eq:Ef}. The ideal gradient $Gf_{i}$ is then calculated with Equation \ref{eq:Gf}.
\begin{equation}\label{eq:Ef}
\footnotesize
  Ef_i = E_i - N_i
\end{equation}
\begin{equation}\label{eq:Gf}
\footnotesize
  Gf_{i} = Ef_{i} - E_{i-1}
\end{equation}
QISMET then employs gradients $Gf_i$ and $G_i$ to govern VQA progress. While this approach can address transient noise with short duration and spike-like magnitude, it proves insufficient for tackling prolonged and non-deterministic noise drift, as shown in Fig.\ref{fig:noise_drift}, leading to deceptive and detrimental iterations during tuning. Furthermore, the noise estimation process requires retrial, re-executing the reference circuit with current circuit configurations, which doubles the computational resources compared to traditional VQA iterations. Further details are discussed in Section \ref{sec:Design}.

%%%%%%%%%%

\section{\name Design} \label{sec:Design}

\name framework overview is depicted in Fig.\ref{fig:overview}. It appends the observable circuits of prime subsets from multiple previous iterations, which serve as references to diligently detect the noise drift in VQA iterations. Noise drift errors are assessed by comparing the energy of the references in previous jobs with their repetitions in the current job. This process enables \name to decide whether to reschedule or accept a particular iteration, thus controlling VQA progression. The orange arrow indicates the action in case of the existence of noise drift: rescheduling of current circuits via the next job. The green line indicates the accepted scenario, proceeding to stage $S_2$ for total energy estimation. The design details of \name are discussed in the following sections, with a focus on three key insights: multi-reference enhancement, computing overhead minimization, and enhanced noise drift detection.

%%%%%%%%%%%%
\subsection{Multi-Reference Enhancement} \label{sec:reference}

The potential for detrimental impacts of noise drift errors is presented in Fig.\ref{fig:noise_drift}, which is seen beyond the statistical robustness offered by executing multiple circuit shots. In such cases, relying solely on a single adjacent iteration for dynamic noise detection, as with \cite{qismet}, can be deceptive to the optimizer and introduce fallacious iterations with inaccurate objective function (energy) estimations during the tuning process. Utilizing such fallacious references to estimated noise errors in subsequent iterations fails to establish a reliable environment for tuners. Instead, the bias accumulates, potentially causing VQA to be far from its target, as shown in Fig.\ref{fig:navi}-d.

\begin{figure}[t]
  \centering
  \includegraphics[width=\linewidth]{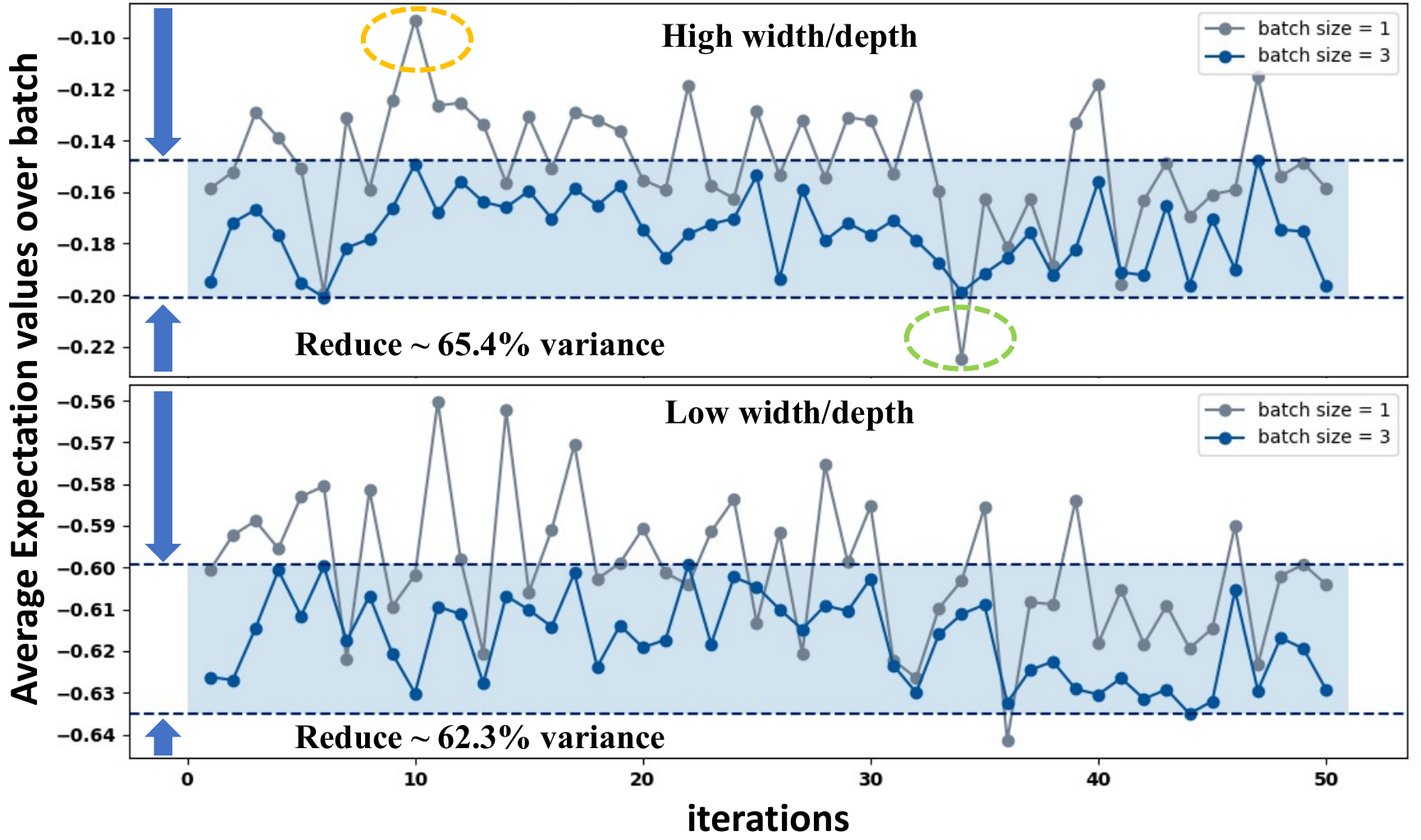}
  % \vspace{-1.5em}
  \vspace{-15pt}
  \caption{The benefits of averaging in reducing deviations. Expectation value data over 50 executions of two circuit batches (blue line: three different circuits; gray line: one circuit) is compared in two different circuit features on IBMQ Lagos.}
  \vspace{-15pt}
  \label{fig:reduce_variation}
\end{figure}

Additionally, deviations in objective function estimation also hinder accurate noise drift detection, particularly when relying on a single reference. Fig.\ref{fig:reduce_variation} presents the circuit expectation values collected over roughly a 24-hour period, featuring two qubit/gate-level characteristics and two circuit batch sizes. The gray line in each plot corresponds to a batch consisting of only one circuit configuration. In the bottom plot, the gray line has an average of about -0.59 and a range value of 0.08. Significant outliers (negative and positive, circled in green and orange, respectively) are observed in the top plot. The gray line in the top plot exhibits an average of around -0.14 and a large range value of 0.13, highlighting the challenges arising from fluctuating results in objective function estimations within noisy landscapes. Such deviations impede accurate detection of the noise drift and result in the termination of the VQA tuning process far from a minimum, such as Fig.\ref{fig:navi}-d.

In contrast to the single-circuit configuration shown in Fig.\ref{fig:reduce_variation}, the blue line in the same figure represents a batch consisting of three different circuits. This approach attenuates a significant portion of the fluctuations in both sub-figures, decreasing the variance by 65.4\% and 62.3\%, respectively. These results indicate that incorporating multiple references improves the reliability of noise detection, thereby mitigating the negative impacts of the aforementioned adverse factors, including fallacious iterations in the reference set and deviations in executing the references. Consequently, the use of multi-references is beneficial for VQA since reliable noise drift detection facilitates maintaining the gradient calculation of VQA in drift-free scenarios.

\subsection{Computing Overhead Minimization} \label{sec:overhead}

Dynamic noise estimation, discussed in Section \ref{sec:detection} requires bundling the circuit configurations corresponding to the current iteration and reference circuit in each VQA iteration. This process doubles the required computing resources compared to a traditional VQA iteration. Section \ref{sec:reference} highlights the benefits of using multiple references, which further increases the computing overhead linearly with the number of reference circuits involved. Additionally, to align the machine-obtained gradient with the ideal gradient, several iterations are skipped when detrimental noise drift occurs. In extreme cases, it may even be necessary to execute the circuit bundle several times to accept a single iteration. Although it is somewhat efficient in mitigating noise drifts, this level of computing overhead is unacceptable for a simple task and is even more severe for long-running applications such as VQA. To eliminate the aforementioned computing overhead, Pauli-term subsetting is proposed.
\begin{figure}[t]
  \centering
  % \vspace{-2em}
  \includegraphics[width=\linewidth]{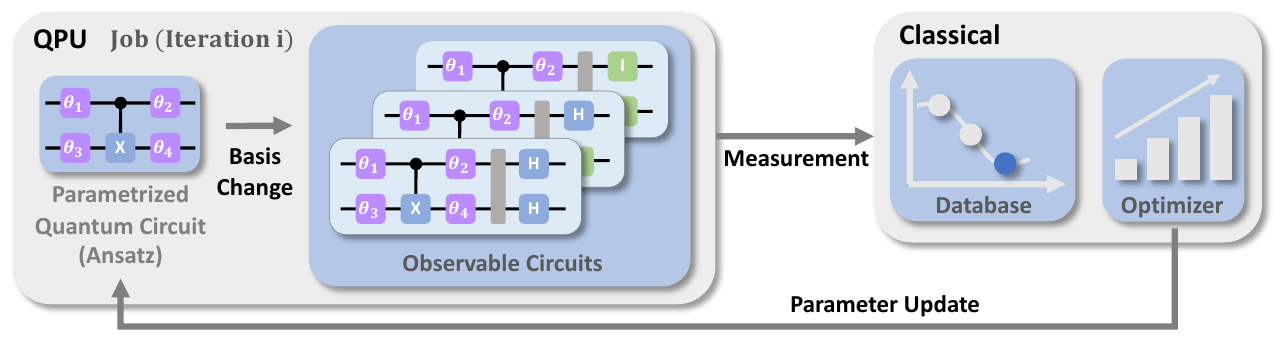}
  \vspace{-15pt}
  \caption{Variational quantum algorithm (VQA), a hybrid quantum-classical algorithm}
  \label{fig:VQA}
  \vspace{-1.5em}
\end{figure}

\begin{figure}[b]
  \centering
    \vspace{-1.5em}
  \includegraphics[width=0.9\linewidth]{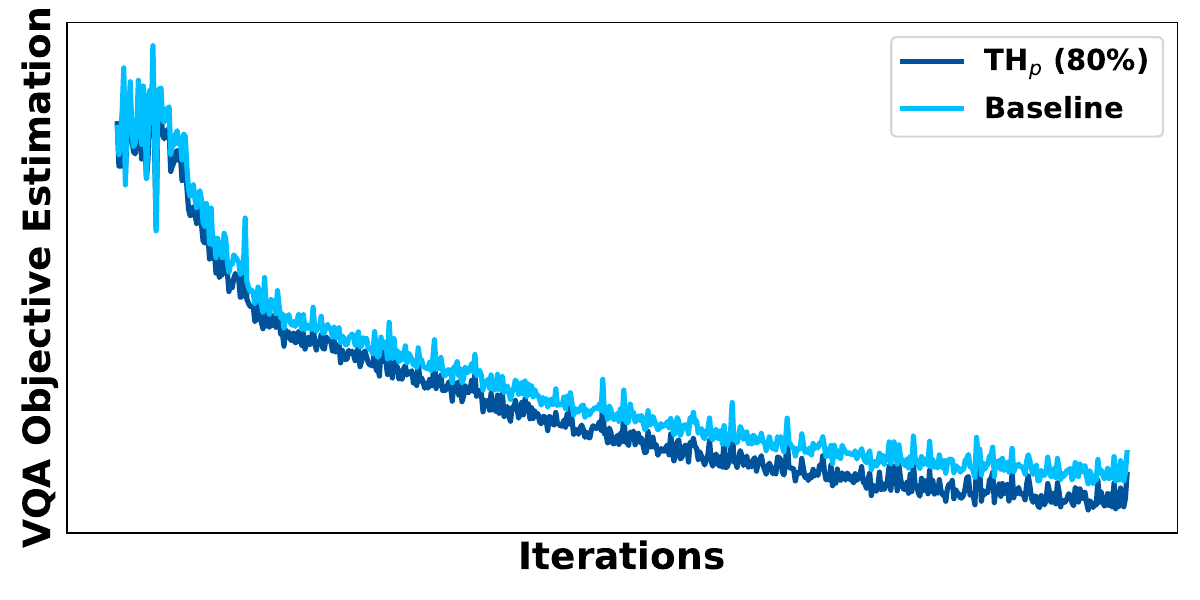}
    \vspace{-15pt}
  \caption{Traditional VQA convergences. The light blue line represents the training trace of the Hamiltonian $\hat{\text{H}}$, while the dark blue line illustrates the convergence of the prime subset with a setting of $\text{TH}_p=80\%$. Prime subset dominates the energy estimation of $\hat{\text{H}}$.
  % The dark blue line is the convergence of prime subset with setting $\text{TH}_p=80\%$.
  }
  \label{fig:prime}
\end{figure}

Equation \ref{eq:Hamiltonian} defines a Hamiltonian $\hat{\text{H}}$, which describes the total energy of a system and serves as the objective function of a problem.
\begin{equation}\label{eq:Hamiltonian}
% \vspace{-0.5em}
\footnotesize
    \hat{\text{H}} = 1.4\cdot \text{XX} + 0.05\cdot \text{ZI} + 0.02\cdot \text{ZX}
\end{equation}
To estimate the total energy (or objective function) for $\hat{\text{H}}$ with a parameterized quantum circuit (ansatz) via VQA, each Pauli term is converted into its corresponding basis to generate observable circuits (OC) for their energy estimation. Subsequently, all terms are aggregated for the total energy of $\hat{\text{H}}$. Fig.\ref{fig:VQA} depicts a VQA iteration of $\hat{\text{H}}$, where the circuits with blue backgrounds are the OCs that are bundled into a ``job'' to be executed for energy estimation \cite{Qiskit}. In other words, even a simple Hamiltonian like $\hat{\text{H}}$ necessitates QPU to execute multiple OCs for a single iteration.

In Equation \ref{eq:Hamiltonian}, the Pauli term ``XX'' holds the largest absolute coefficient value, which accounts for over 95\% of all the coefficients of Pauli terms. This indicates that the OC corresponding to the Pauli term ``XX'' (the circuit in Fig.\ref{fig:VQA}, with each qubit followed by a Hadamard gate) significantly impacts the energy estimation of $\hat{\text{H}}$, as shown in Fig.\ref{fig:prime}. Hence, we introduce a new term ``prime subset'' to refer to such OCs that are deemed to dominate the energy estimation of a Hamiltonian. Specifically, the OC with an absolute coefficient exceeding a specified threshold value $\text{TH}_p$ relative to the sum of absolute coefficients for all Pauli terms. The setting of threshold value $\text{TH}_p$ is discussed further in Section \ref{sec:eva_th}. Note: The prime subset is not limited to a single circuit but rather is a collection of circuits. Additionally, we define a ``minor subset'', which represents the remaining OCs.

Based on the Pauli-subsetting design, \name actively takes the following steps in each iteration to alleviate computing overhead without sacrificing the reliability of noise drift detection: (i) rearranging the execution order of OCs corresponding to the current iteration by partitioning them into two stages (prime subset in $S_1$, minor subset in $S_2$); (ii) executing the prime subset of reference circuits for noise drift detection in $S_1$, instead of all the references' observable circuits; (iii) proceeding to $S_2$ only if the current iteration passes the noise drift detection.

% \subsection{\itname Implementation}
% \subsection{\textbf{Enhanced \itname Noise Drift Detection}}

\subsection{\itname Noise Drift Detection}

As discussed in Section \ref{sec:reference} and Section \ref{sec:overhead}, \name incorporates multi-references and deploys Pauli-term subsetting to improve the reliability of noise drift detection with less execution time in $S_1$ stage (Note: The adjacent previous iterations are usually suitable choices for reference since they typically have similar objective function estimations to the circuit configuration in the current iteration, and the adjacent iterations generally have closer noise landscapes compared to other iterations). The noise drift error estimation in Equation \ref{eq:D} is enhanced to incorporate $K$ reference circuits from iteration $\{(i-n) \mid n \in [1..K]\}$:
\begin{equation}\label{eq:up_D}
\footnotesize
    D_i = \sum^K_{n=1}c_{i-n}[Er_{i-n}(P) - E_{i-n}(P)] \; s.t.\; \sum^K_{n=1}c_n = 1
\end{equation}
where $(P)$ denotes the prime subset, $K$ represents the incorporated reference number, which is discussed next in Section \ref{sec:eva_ref}, and $c_{i-n}$ is the proportion factor corresponding to the reference circuits from the previous iteration $(i-n)$. $c_{i-n}$ is fixed at $\frac{1}{K}$ to ensure the sum of all factors equals 1 (Note: dynamic factors could potentially enhance benefits even further, but this is beyond the scope of our current study). 

With the design of $K$ reference circuits, \name provides a more faithful estimation of the drift-free energy $Ef_i(P)$ and the corresponding drift-free gradient $Gf_{i}$. The calculation process is enhanced using the updated Equations \ref{eq:up_Ef} and \ref{eq:up_Gf}, respectively:
\begin{equation} \label{eq:up_Ef}
\footnotesize
  Ef_i(P) = E_i(P) - D_i
\end{equation}
\begin{equation} \label{eq:up_Gf}
\footnotesize
  Gf_{i} = Ef_{i}(P) - \sum^K_{n=1}c_{i-n}\cdot E_{i-n}(P)
\end{equation}
The multi-references based machine-obtained $G_i$ in Equation \ref{eq:G} is described below:
\begin{equation} \label{eq:up_G}
\footnotesize
% \vspace{-0.5em}
  G_{i} = E_i(P) - \sum^K_{n=1}c_{i-n}\cdot E_{i-n}(P) 
\end{equation}

\begin{figure}[t]
  \centering
  % \vspace{-15pt}
  \includegraphics[width=\linewidth]{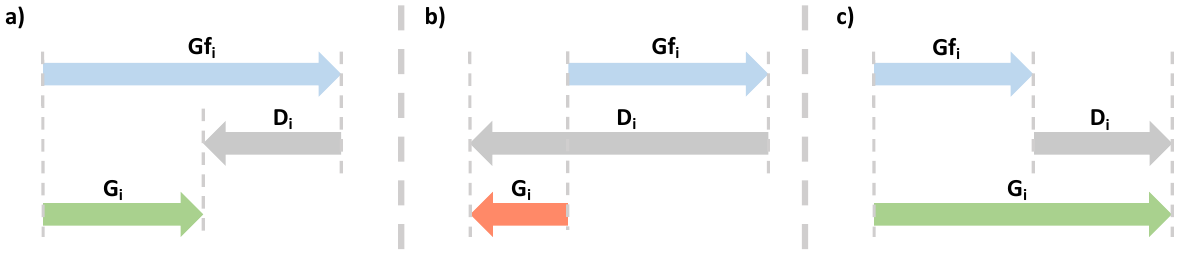}
  \caption{\name control policy. a) and c): Machine-obtained and drift-free gradients have the same direction, thus acceptable; b): Noise drift flips the machine-obtained direction, thus iteration is rescheduled.}
  \label{fig:condi}
  \vspace{-15pt}
\end{figure}

To handle various intractable noise drift, \name employs an intelligent control policy to diligently guide VQA, ensuring its training remains in a mild landscape. The aforementioned gradients $Gf_i$ and $G_i$ are utilized to make informed decisions, which control VQA progress by accepting or rescheduling particular iterations. The underlying principle is to accept VQA iterations only if the direction of $Gf_i$ (gradient obtained from \name noise drift detection) coincides with the direction of $G_{i}$ (gradient observed by the VQA tuner based on machine energy estimations). Fig.\ref{fig:condi}-a, c describe the acceptance scenarios. This precautionary approach ensures that the entire VQA tuning process takes place under the same or similar landscapes, enabling the tuner to steadily and reliably approach its objective without deviating from the target due to the negative impact of noise drift.

For the scenario in Fig.\ref{fig:condi}-b, in response, \name skips the result in the current iteration due to noise drift errors and reschedules the job. As a result, identical circuit configurations are repeated in the next iteration. Note that noise drift effects can persist for extended periods of time. Therefore, the skipping and repetition may span multiple jobs but are limited by a max-out $\sigma$. Once max-out is reached, \name deems that the landscape has completely changed, rendering the references' energy in previous jobs inapplicable to the current landscape. In this situation, \name updates the reference energy (used only for detection) and detects the noise drift based on the updated reference energy. Algorithm \ref{alg} outlines the process.

\begin{algorithm}[b]
\caption{\name Workflow} \label{alg}
\begin{algorithmic}
\footnotesize
\Require execute $N$ iterations
% \Ensure $min[\langle\psi({\overrightarrow{\theta}}) |\hat{H}| \psi({\overrightarrow{\theta}})\rangle]$
% \STATE $ \hat{H} \gets \mathcal{C}_p + \mathcal{C}_m$
% \Ensure $n \in [1..3]$
\State $i=0$, $i_r=0$
\While{$i \neq N$}
\State $\{E_i(P), Er_{i-n}(P)\}\gets QPU(P_i, P_{i-n}) \mid n \in [1..K]$ 
\For{$n \in [1..K]$}
    \State $E_{i-n}(P) \gets \text{Database} $ 
\EndFor

\State $D_i \gets \sum^K_{n=1}c_{i-n}[Er_{i-n}(P) - E_{i-n}(P)]$
\State $Ef_i \gets E_i(P) - D_i$
\State $Gf_{i} \gets Ef_{i}(P) - \sum^K_{n=1}c_{i-n}\cdot E_{i-n}(P)$
\State $G_{i} \gets E_i(P) - \sum^K_{n=1}c_{i-n}\cdot E_{i-n}(P)$

\If{$ G_{i}\cdot Gf_{i} > 0$}
    \State $E_i(M) \gets QPU(M_i)$ 
    \State $i \gets i + 1$
    \State $\text{Database} \gets E_i(P)+E_i(M)$
\Else
    \State $i_r \gets i_r + 1$
    \If{$i_r == \sigma$}
    \State $\text{Database} \gets \{Er_{i-n}(P)\}  \mid n \in [1..K]$
    \State $\text{Database} \gets E_i(P)+E_i(M)$
    \State $i_r \gets 0$
    \EndIf
\EndIf

\EndWhile
\end{algorithmic}
\end{algorithm}
\begin{figure*}[ht]
  \centering
  \includegraphics[width=0.95\linewidth]{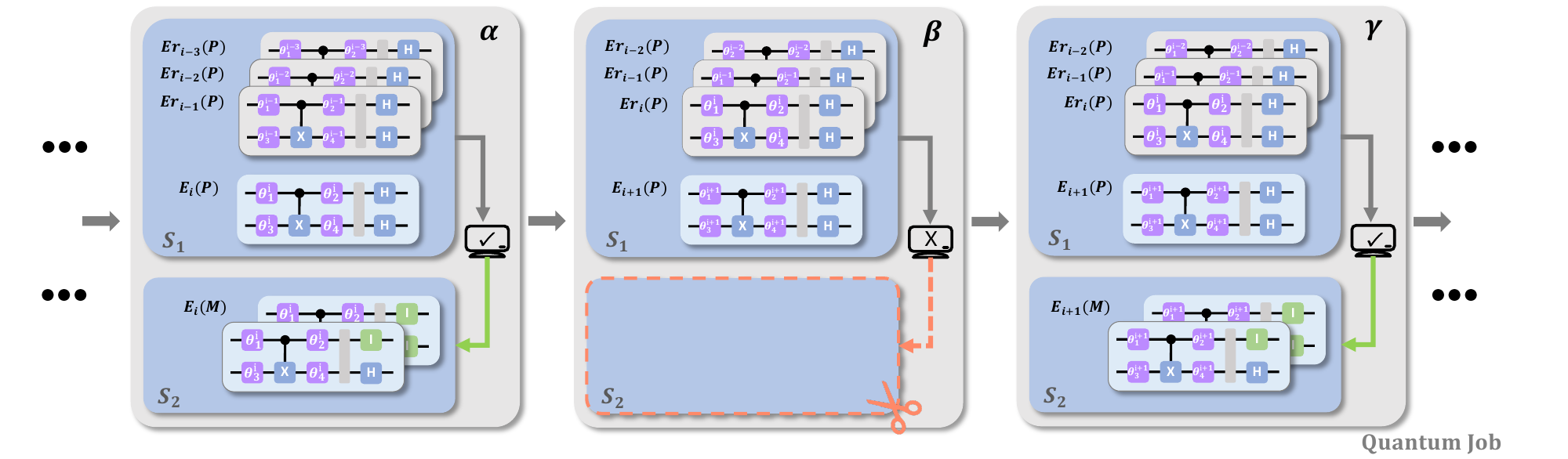}
  \vspace{-12pt}
  \caption{Multiple VQA ``jobs'' (gray box) run on a QPU. Each job has two execution stages. $S_1$ consists of the prime subset corresponding to the current VQA iteration (circuits with blue background) and the prime subset of reference circuits appended by \name (circuits with gray background). $S_2$ is determined by \name (as check mark $\checkmark$ or X) and executes the minor subset corresponding to the current VQA iteration. VQA progresses when both $S_1$ and $S_2$ have completed.}
  \label{fig:btjobs}
  \vspace{-15 pt}
\end{figure*}

\subsection{Functionality Across Iterations}
\name workflow with three references over several continuous VQA iterations is illustrated in Fig.\ref{fig:btjobs}.
In $S_1$ stage of job $\alpha$, QPU executes the circuits consisting of the prime subset corresponding to iteration $(i)$, depicted as the inner box with blue background, and the configuration of prime subsets corresponding to iteration $\{(i-n) \mid n \in [1..3]\}$, depicted with light gray background. Then, the input of the noise drift detection includes the above executed results $E_i(P)$ and $\{Er_{i-n}(P) \mid n \in [1..3] \}$ along with reference prime subset energy from previous jobs $\{E_{i-n}(P) \mid n \in [1..3] \}$ (not shown in the figure). In job $\alpha$, \name indicates the impact of noise drift on iteration $(i)$ is acceptable (check mark). Therefore, job $\alpha$ proceeds to $S_2$ stage and the minor subset corresponding to iteration $(i)$ is scheduled to execute for the total energy estimation of iteration $(i)$. VQA thus progresses to iteration $(i+1)$.\\
Job $\beta$ is the adjacent job right after job $\alpha$ to compute the circuits of iteration $(i+1)$. The prime subset of iteration $(i)$ replaces the prime subset of iteration $(i-3)$ as one of the reference circuits for this job. Other executed circuits in stage $S_1$ are shown in the figure. Once stage $S_1$ of $\beta$ is executed, the following information is utilized for noise drift detection:\\
a) prime subset energy corresponding to iteration $(i+1)$;\\
b) $\{E_{i-n}(P) \mid n \in [0..2]\}$, prime subset energy of reference circuits from the previous jobs;\\
c) $\{Er_{i-n}(P) \mid n \in [0..2] \}$, the repetitions of reference circuits in the current job $\beta$.

In this example, \name detects that the noise drift error is out of range and derails VQA convergence (indicated by an X in the figure). Therefore, iteration $(i+1)$ is rescheduled, and the outcomes of job $\beta$ are discarded. The figure shows the termination of job $\beta$ ($S_2$ is not executed) and a subsequent repetition of circuits via job $\gamma$.\\
Once stage $S_1$ of job $\gamma$ is completed, \name checks again and determines whether the noise drift present in job $\beta$ that still exists in job $\gamma$. 
In this job, \name observes that the noise drift has passed or deems its impact to be within an acceptable range. The $S_2$ stage of job $\gamma$ is processed and contributes further to the progression of VQA. By taking above steps in each VQA iteration, \name provides a stable landscape for the gradient computation of the tuning process.

\section{Evaluation}

\subsection{Evaluation Methodology}\label{sec:config}
\subsubsection{\textbf{General Infrastructure}} 
\name is a software optimization approach implemented using the Qiskit Runtime Python package \cite{Qiskit}, which allows for running quantum programs entirely on the IBM Cloud and executing workloads on quantum systems at scale. \name is broadly applicable across all VQA applications and can be integrated into any VQA classical optimizer to enhance noise drift resilience. In this work, we restrict the evaluation to eight qubits considering the machine limitations on circuit metrics (depth, width). The evaluation deploys SPSA as the main classical optimizer and primarily focuses on VQE, one of the VQA applications (introduced in Section \ref{sec:VQA}). The evaluations encompass four Hamiltonians, four different ansatzes, and six QPUs from IBM.

\subsubsection{\textbf{Benchmarks}}
The primary Hamiltonian evaluated is the potential energy of the helium hydrogen ion over bond lengths of 1.7 $\dot{A}$, with additional evaluations for the potential energy of the hydrogen fluoride molecule, the lithium hydride molecule, and the hydrogen molecule. The molecules are labeled with a superscript ``c'', indicating that the number of terms in their Hamiltonians has been reduced using the reduction method from \cite{compress}. This method is capable of compressing large fermionic Hamiltonians into several qubits, enabling more efficient calculations and analysis. The hardware-efficient SU2 \cite{SU2} and RA \cite{RA} ansatz are used in the experiments, varying the block repetitions between 2, 6, and 10 to change the number of parameters in the ansatz. The selected QPUs are Kolkata (27 qubits), Toronto (27 qubits), Montreal (27 qubits), Perth (7 qubits), Jakarta (7 qubits), and Lagos (7 qubits). All the machine details can be found at IBMQ's website \cite{ibmq}. 

Despite the convenience of accessing IBM's quantum machines through the cloud and facilitating the workload with Qiskit Runtime, the limited access to quantum machines still prevents a holistic evaluation of our proposal. To enable fine-granularity noise drift evaluation, we combine the noise trace generated by the IBM machine Toronto with the enhanced noise traces upon the model from \cite{qismet}. This enhanced model captures the noise drift effects in each iteration and normalizes these effects to the magnitude of VQA estimations, allowing the simulations to exactly mimic observed noise drift errors. 

\subsubsection{\textbf{Baselines}} The following schemes are assessed in several comparative evaluations:
\begin{itemize}
  \item \name: Setting the number of optimal reference circuits to three (two for HeH$^+$ ion) and the prime threshold $\text{TH}_p$ to 80\%.
  
  \item Baseline: Traditional variational quantum eigensolver with deploying SPSA for optimization.
  
  \item QISMET: Transient noise mitigation approach \cite{qismet} with tuning the threshold to 10\% to provide optimal performance.

\end{itemize}

% (iv) \textbf{\name-S}: \name with setting the number of optimal reference circuits to one.\\
% (v) \textbf{\name-A}: \name with considering all the OCs as the reference circuits and setting the number of optimal reference circuits to three.\\
% \textbf{\name-S}: \name without partitioning execution into $S_1$ and $S_2$.

%%%%%%%%%%%%
\subsection{Evaluation Results}
\subsubsection{\textbf{Evaluation on QPUs}}
The evaluation of \name, QISMET, and Baseline for VQA energy estimation is conducted on IBMQ machines to solve the VQA energy estimation problem for a four qubits HeH$^+$ ion, with a fixed number of 350 iterations. The experiments are conducted synchronously to ensure temporal adjacency between the iterations of \name and other schemes.
\begin{figure}[b]
  \centering
  \vspace{-15pt}
  \includegraphics[width=0.9\linewidth]{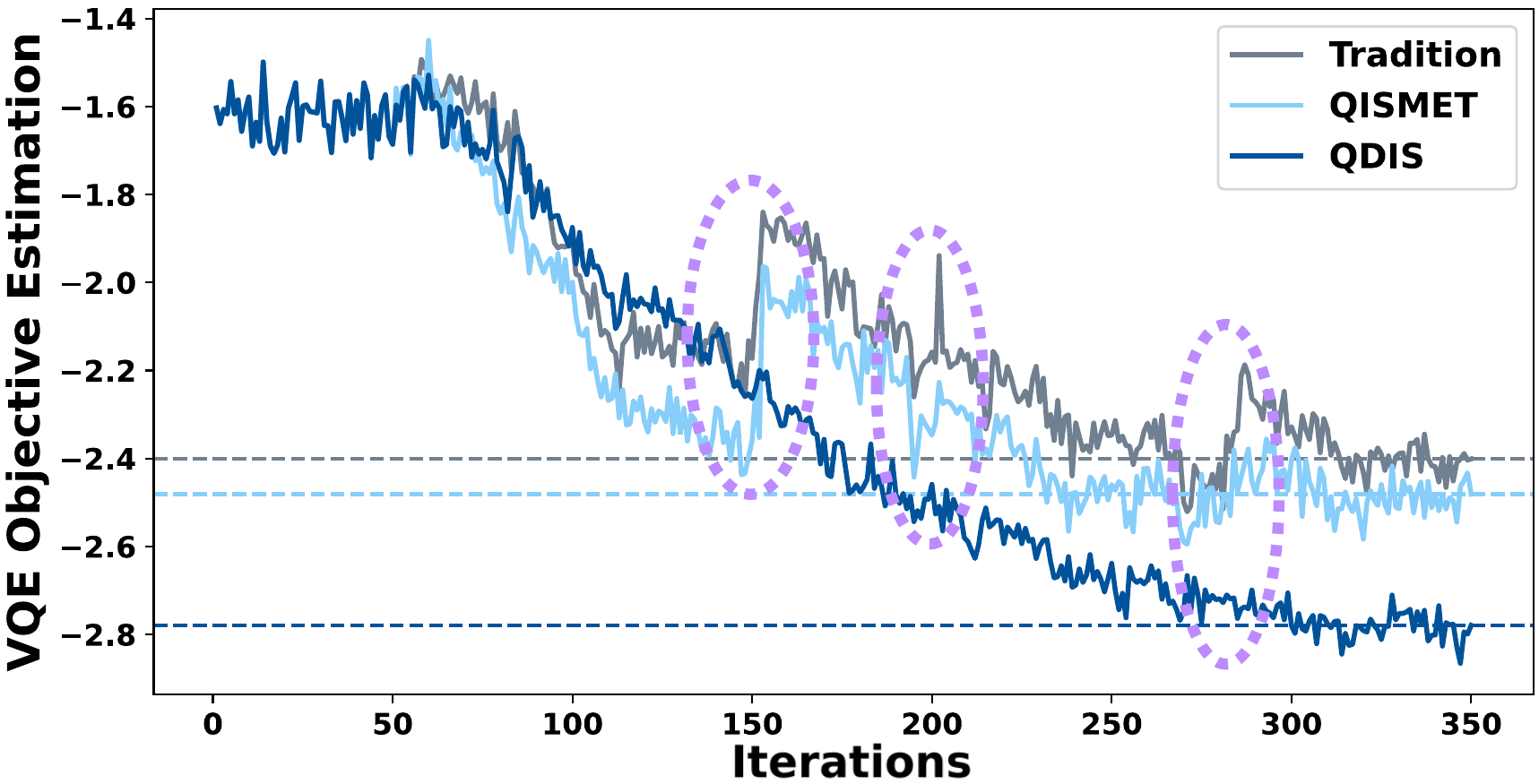}
  \vspace{-8pt}
  \caption{\name benefits for a HeH$^+$ VQA application on IBMQ Montreal with several high fluctuating noise regions (circled). Sharp noise drift errors are avoided by \name.}
  \label{fig:Montreal}

\end{figure}

The results in Fig.\ref{fig:Montreal} compare \name against other schemes on IBMQ Montreal device. Two periods of significant fluctuations, which heavily derail both QISMET and Baseline, are highlighted. QISMET only eliminates the noise instance in the second period and fails to avoid the first serious turbulence due to the variation in perceived gradient estimates. In contrast, benefiting from multi-reference circuits, \name predominantly bypasses both turbulent periods ensuring steady progress, ultimately achieving improvements of 59.8\% and 44.9\% over Baseline and QISMET, respectively.

\begin{figure}[t]
  \centering
% \vspace{-1.5em}
  \includegraphics[width=0.9\linewidth]{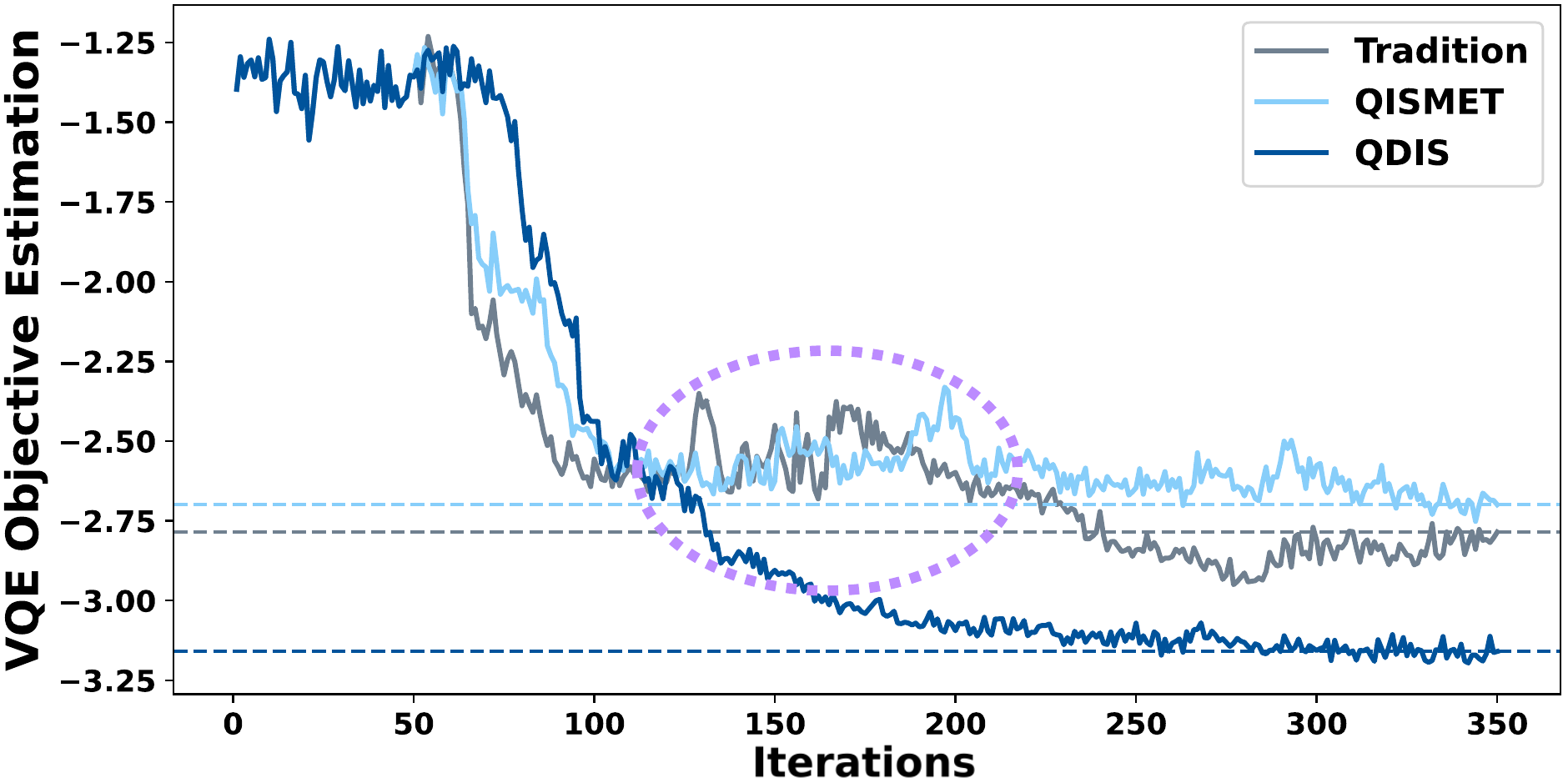}
  \vspace{-9pt}
  \caption{A HeH$^+$ VQA application on IBMQ Toronto with an instance of noise drift error spanning multiple iterations (circled). This phase of noise is bypassed by \name, thus improving convergence of application.}
  \label{fig:Toronto}
  % \vspace{-5pt}
\end{figure}

Fig.\ref{fig:Toronto} shows a comparison of \name against Baseline and QISMET on IBMQ Toronto. The noise drift errors persist for an extended period (from 100 to 170 iterations), resulting in multiple instances of noise drift errors with moderate or tiny magnitudes. The tuning of QISMET and Baseline stagnates during this period. Although \name is also impacted for several early iterations, it quickly recovers and continues its navigation to the target, improving fidelity over Baseline and QISMET by 29.0\% and 37.4\%, respectively.

\begin{figure}[t]
  \centering
  \vspace{-10pt}
  \includegraphics[width=0.9\linewidth]{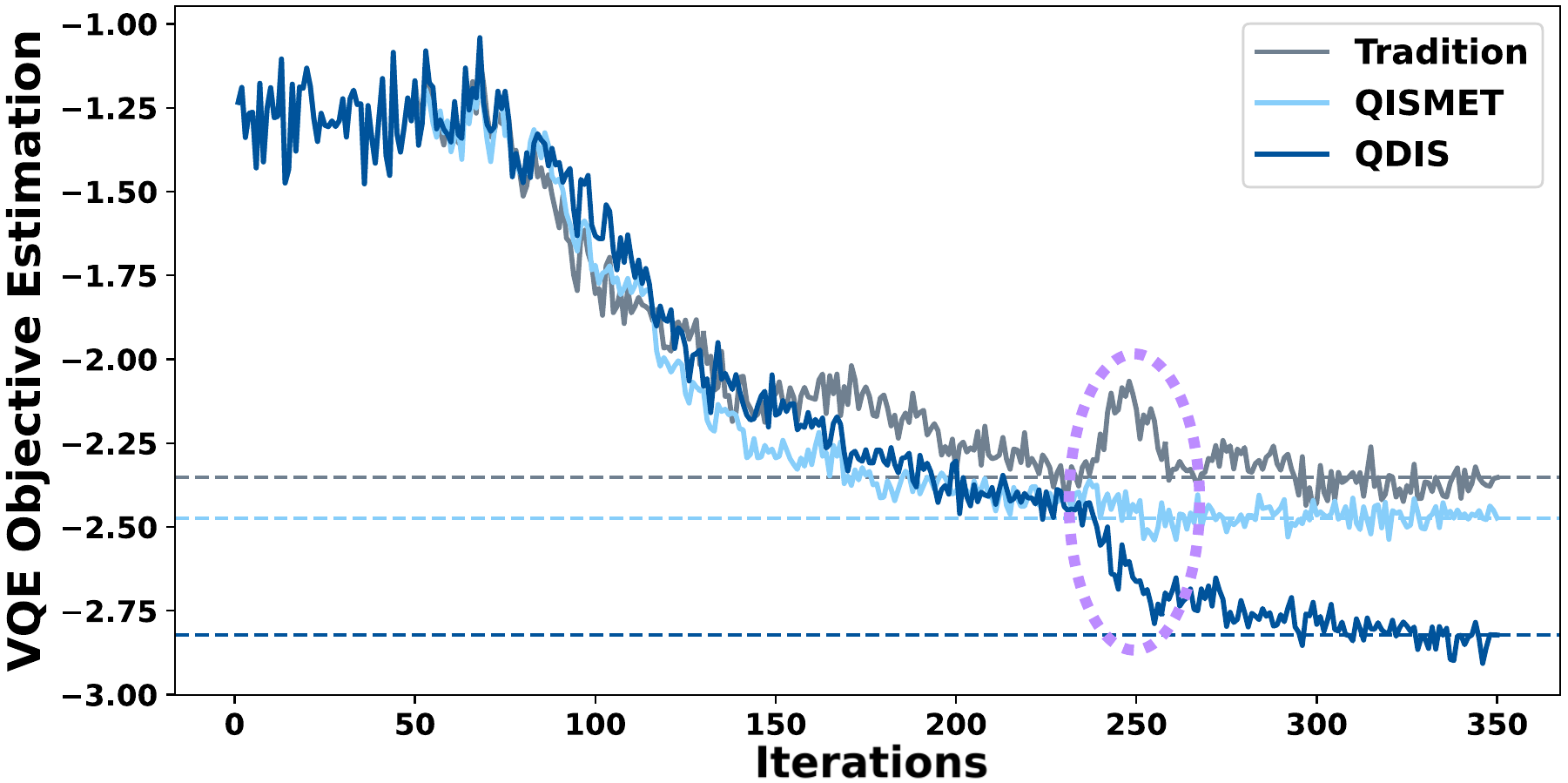}
  \vspace{-9pt}
  \caption{A HeH$^+$ VQA application on IBMQ Jakarta with an inconspicuous instance of noise drift error (circled). The convergence of application benefits from \name by skipping this malicious instance of noise.}
  \label{fig:Jakarta}
  % \vspace{-2em}
  \vspace{-1.5em}
\end{figure}
\begin{figure}[b]
  \centering
  \vspace{-20pt}
  \includegraphics[width=\linewidth]{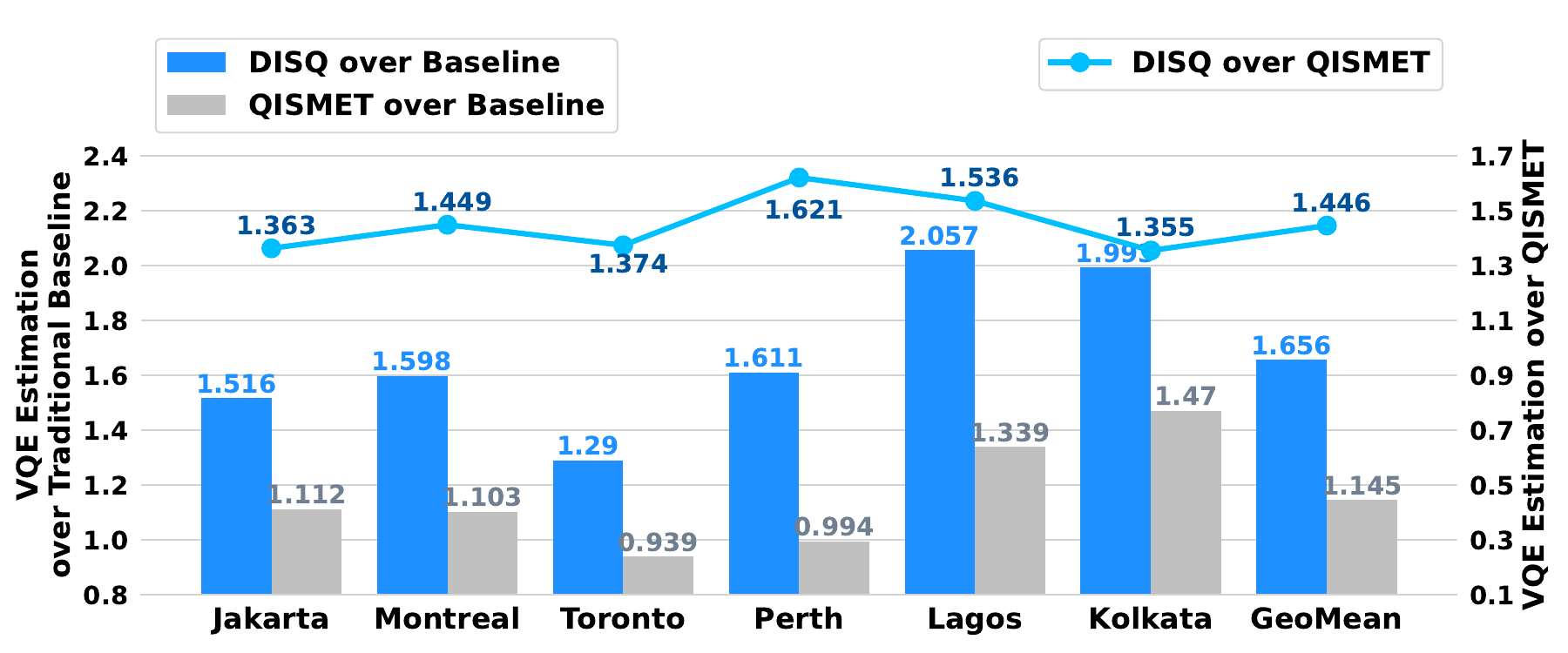}
  \vspace{-20pt}
  \caption{\name's benefits for a HeH$^+$ VQA application on six IBMQ machines}
  \label{fig:machine_result}
  % \vspace{-1.5em}
\end{figure}
In the evaluation on IBMQ Jakarta shown in Fig.\ref{fig:Jakarta}, the machine behavior is mostly smooth except for one instance of an inconspicuous noise drift error, which is highlighted. While this period does not seem to cause any severely detrimental impact, QISMET is deceived by a mediocre estimation and languishes at a local optimum. \name heuristically identifies this camouflaged noise drift period and is able to further progress toward the global optimum target with benefits of 51.6\% and 36.3\% over Baseline and QISMET, respectively.

The benefits of \name over Baseline and QISMET on six different IBMQ machines are shown in Fig.\ref{fig:machine_result}. The primary vertical axis shows the improvements in measured VQA expectations over Baseline, while the secondary vertical axis shows the improvement over QISMET. Across all the machines, \name consistently improves the expectation values over Baseline by 1.29-2.057$\times$, with a mean improvement of 1.67$\times$. Furthermore, \name boosts the fidelity by an average of $45\%$ (up to $62.1\%$ and by at least $35.5\%$) over QISMET. The above improvements are achieved over 350 iterations due to access constraints. It is expected that benefits increase with more iterations since this would provide higher potential for more noise drift errors.

%%%%%%%%%%%%
\subsubsection{\textbf{Evaluation with Multiple Benchmarks}} 
\begin{figure}[b]
  \centering
  \vspace{-20pt}
  \includegraphics[width=\linewidth]{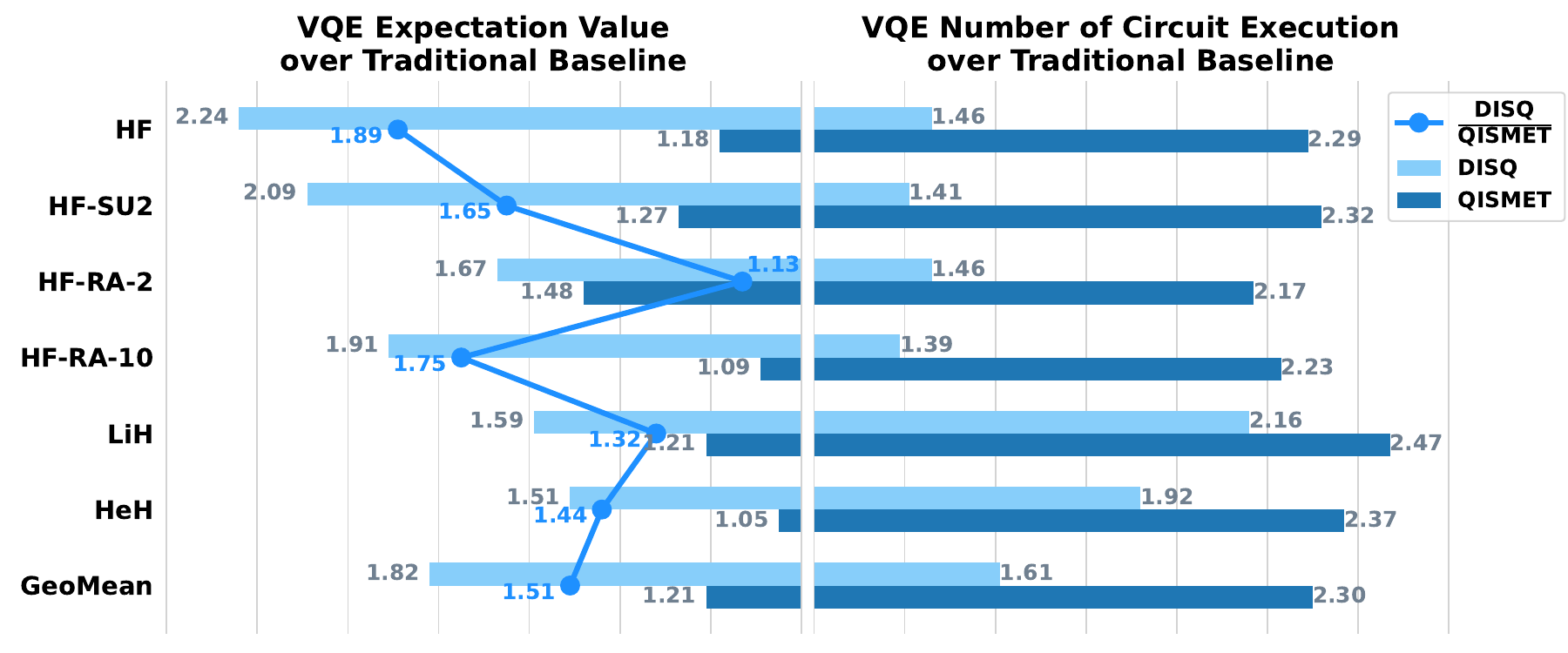}
  \vspace{-20pt}
  \caption{\name provides consistent benefits outperforming the baseline in six different applications.}
  \label{fig:sim_results}
  % \vspace{-1.5em}
\end{figure}

A thorough comparison is conducted with QISMET and Baseline to comprehensively evaluate the performance of \name across the six different applications listed in Table \ref{tab:apps}. The evaluation is conducted over 1000 simulation iterations using different molecules and ansatzes to explore the adaptability of \name. The total number of OC for each molecule is listed under the ``\# Observable'' column and the values in the parentheses are the number of OC in its prime subset. Expectation value comparisons are presented on the left side of Fig.\ref{fig:sim_results}. \name consistently outperforms Baseline, achieving improvements ranging from 1.51$\times$ to 2.24$\times$. Moreover, \name improves the fidelity by up to 1.89$\times$ over QISMET. Multi-reference circuit scheduling employed in \name showcases its superior performance under convoluted, noisy traces across various molecules and ansatzes, which is consistent with the discussion presented in section \ref{sec:reference}. 
\begin{table}[t]
  \centering
  \scriptsize
  \caption{APPLICATION INFORMATION SUMMARY}
  % \vspace{-0.5em}
  \label{tab:apps}
  
  \begin{tabular}{cccccl}
    \toprule
     Application & Molecule & Qubits & Ansatz & Layer Reps & \# Observable \\
    \midrule
    HF & HF  & 8 & RA  & 6  & 76 (8)\\
    HF-SU2 & HF  & 8 & SU2 & 6  & 76 (8)\\
    HF-RA-2 & HF  & 8 & RA  & 2  & 76 (8)\\
    HF-RA-10 & HF  & 8 & RA  & 10 & 76 (8)\\
    LiH & LiH & 6 & RA  & 6  & 13 (4)\\
    HeH & HeH$^+$ & 4 & RA  & 6  & 4  (1)\\
  \bottomrule
\end{tabular}
\vspace{-15pt}
\end{table}

The number of circuit executions in each scheme enables an intuitive comparison of computation overhead, depicted on the right side of Fig.\ref{fig:sim_results}. The leverage of Pauli-term subsetting not only allows \name to schedule more reference circuits in noise detection, but also boosts its computing speed. \name actively reduces executed circuit cost by up to 39.2\% with a mean of 23.5\% across the application, and boosts noise drift detection speed by an average of 2.07$\times$ over QISMET. 
% computational overhead savings 

%%%%%%%%%%%%
\subsubsection{\textbf{Molecule Estimation Evaluation}}

\begin{figure}[t]
  \centering
  % \vspace{-2em}
  \includegraphics[width=0.9\linewidth]{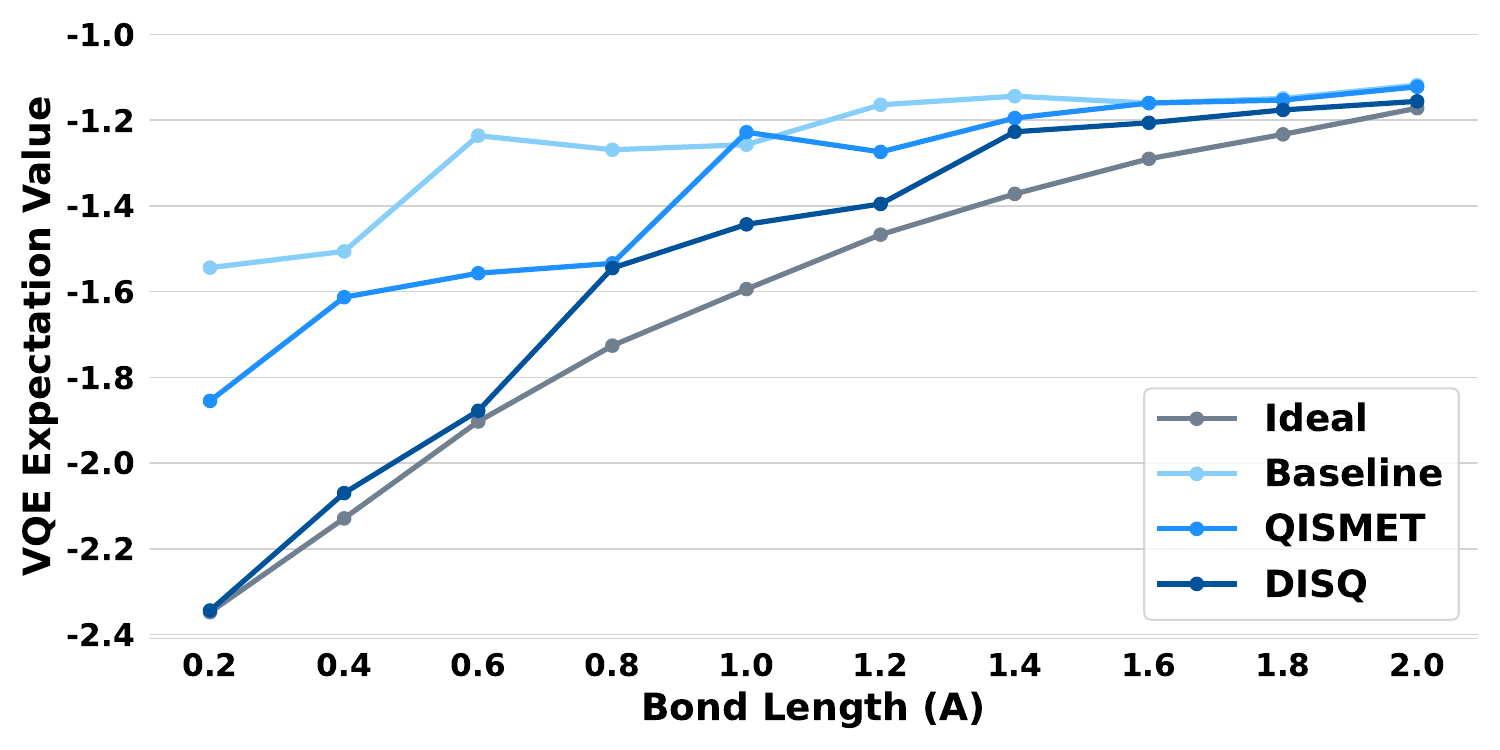}
  \vspace{-15pt}
    \caption{\name benefits in energy estimation for H$_2$ over different bond lengths.}
  \vspace{-15pt}
  \label{fig:VQE}
\end{figure}

VQE is extensively used in molecule chemistry to estimate the energy value of a molecule in a specific geometry, where energy variation indicates the chemical reaction rates for the molecule. The geometry of a molecule typically varies with different bond lengths, resulting in a multitude of Hamiltonians \cite{VQE}. The energy of molecules is calculated from the expectation value of their Hamiltonians. However, noise drift errors affect the expectation values differently for different bond lengths, leading to skewed energy differences. Fig.\ref{fig:VQE} depicts the potential expectation for the $H_2$ molecule over 10 different H-H bond lengths, each corresponding to a unique Hamiltonian and VQE experiment. The gray line represents the noise-free scenario, while \name closely models the noise-free trace by accurately estimating the expectation for each bond length. However, QISMET and Baseline deviate from the ideal scenario as the bond length descends, especially at lower bond lengths where the noise drift errors have a substantial impact. At a bond length of 0.8 $\dot{A}$, the step-wise estimation difference relative to the previous length of 0.6 $\dot{A}$ has opposite gradients when comparing Baseline and Ideal, which results in misleading indications in the chemical reaction rates for Baseline.

\begin{figure}[b]
  \centering
  \vspace{-20pt}
  \includegraphics[width=\linewidth]{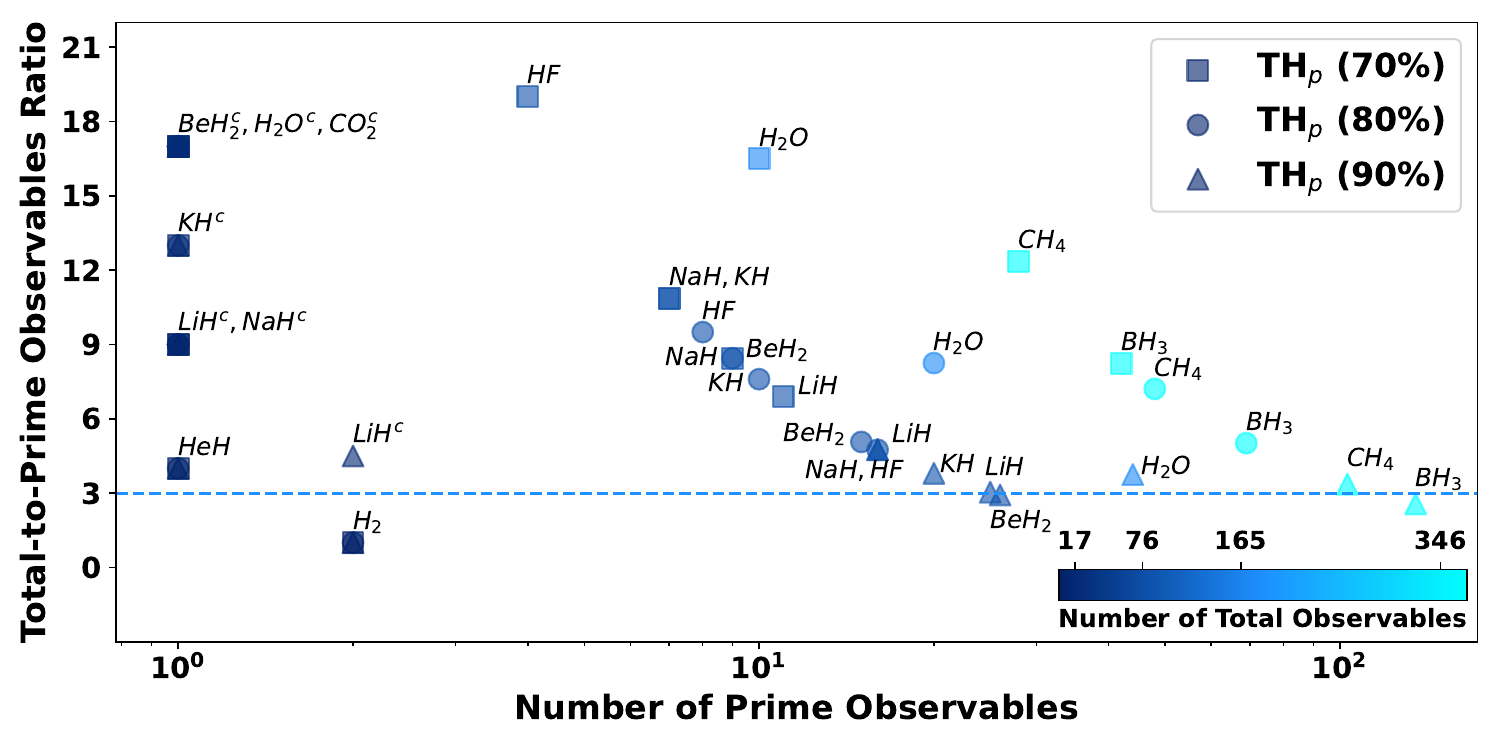}
  \vspace{-20pt}
  \caption{Statistics of prime subset and the ratio of total-to-prime across different molecules with different $\text{TH}_p$ values.}
  \label{fig:scatter}
  % \vspace{-1.5em}
\end{figure}

%%%%%%%%%%%%
\subsection{Analysis}
\subsubsection{\textbf{Sweeping over \itname Prime Threshold}}\label{sec:eva_th}

In section \ref{sec:overhead}, we discuss the use of a prime threshold $\text{TH}_p$ to delineate the criteria for Pauli-term subsetting. To identify an appropriate threshold for the prime subset, we conduct a statistical analysis of the observable circuit (OC) number in different molecules with different $\text{TH}_p$. Our findings reveal that the energy of most molecules heavily relies on a few OCs, aligning with our prime subset design and highlighting the substantial potential of \name to conserve computational resources.

The statistical results from our analysis are summarized in Fig.\ref{fig:scatter}, which depicts evaluated molecule configurations in terms of their OC numbers in the prime subset and total-to-prime ratio (the ratio of the total number of OCs to the number of OCs in the prime subset). Each point is color-coded by OC number and shaped by the applied $\text{TH}_p$. The y-axis represents the total-to-prime ratio, with higher points indicating greater potential for \name to reduce computing overhead. The x-axis (log scale) represents the OC number in the prime subset. The molecules, denoted by the superscript $c$, represent compressed molecules, as introduced in Section \ref{sec:config}. 

\begin{figure}[t]
  \centering
  % \vspace{-1.5em}
  % \includegraphics[width=0.9\linewidth]{Figures/threshold.pdf}
  \includegraphics[width=\linewidth]{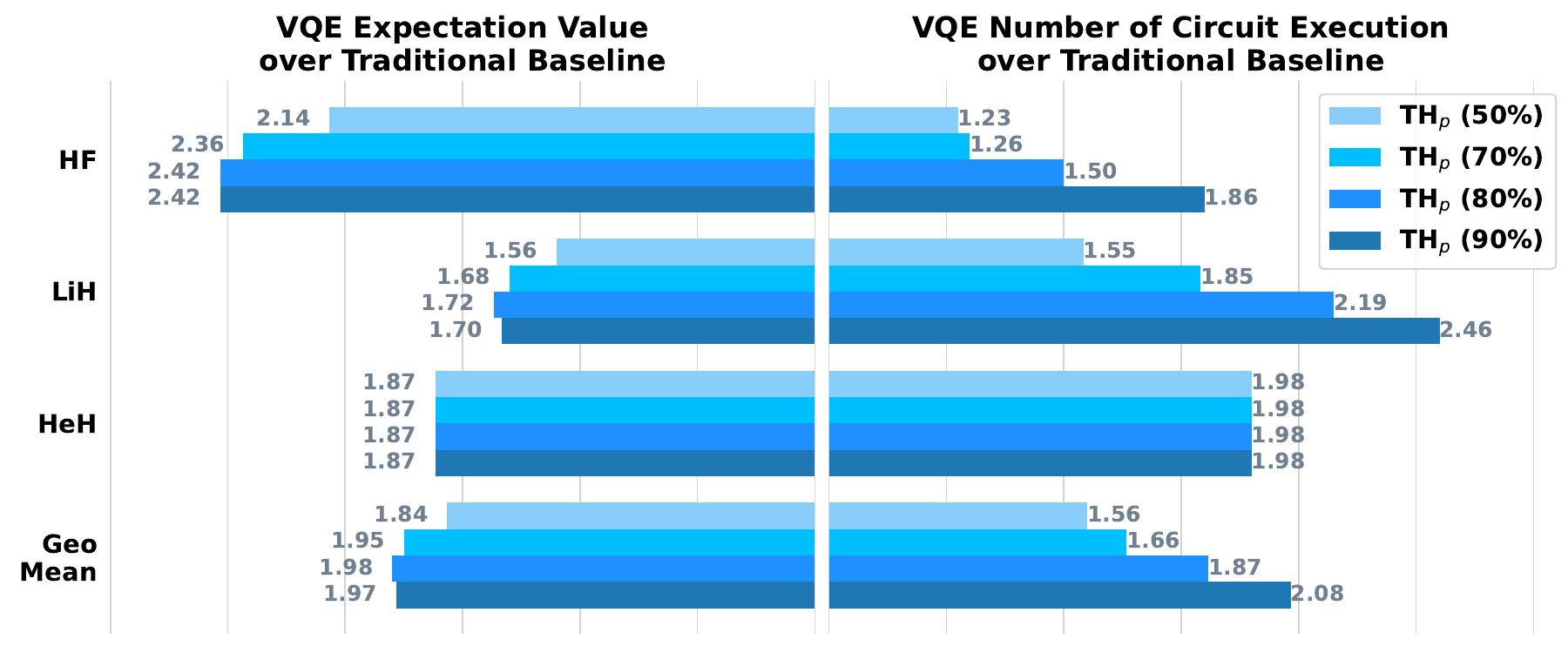}
  \vspace{-20pt}
  \caption{Simulation evaluation with varying \name prime threshold values $\text{TH}_p$. HeH$^+$ and LiH molecules exhibit the same number of OC at $\text{TH}_p = 60$ and $\text{TH}_p = 70$, while HeH$^+$ maintains a consistent OC count across these thresholds.}
  \label{fig:threshold}
  % \vspace{-10pt}
  \vspace{-20pt}
\end{figure}

\begin{figure}[b]
  \centering
  \vspace{-20pt}
  \includegraphics[width=\linewidth]{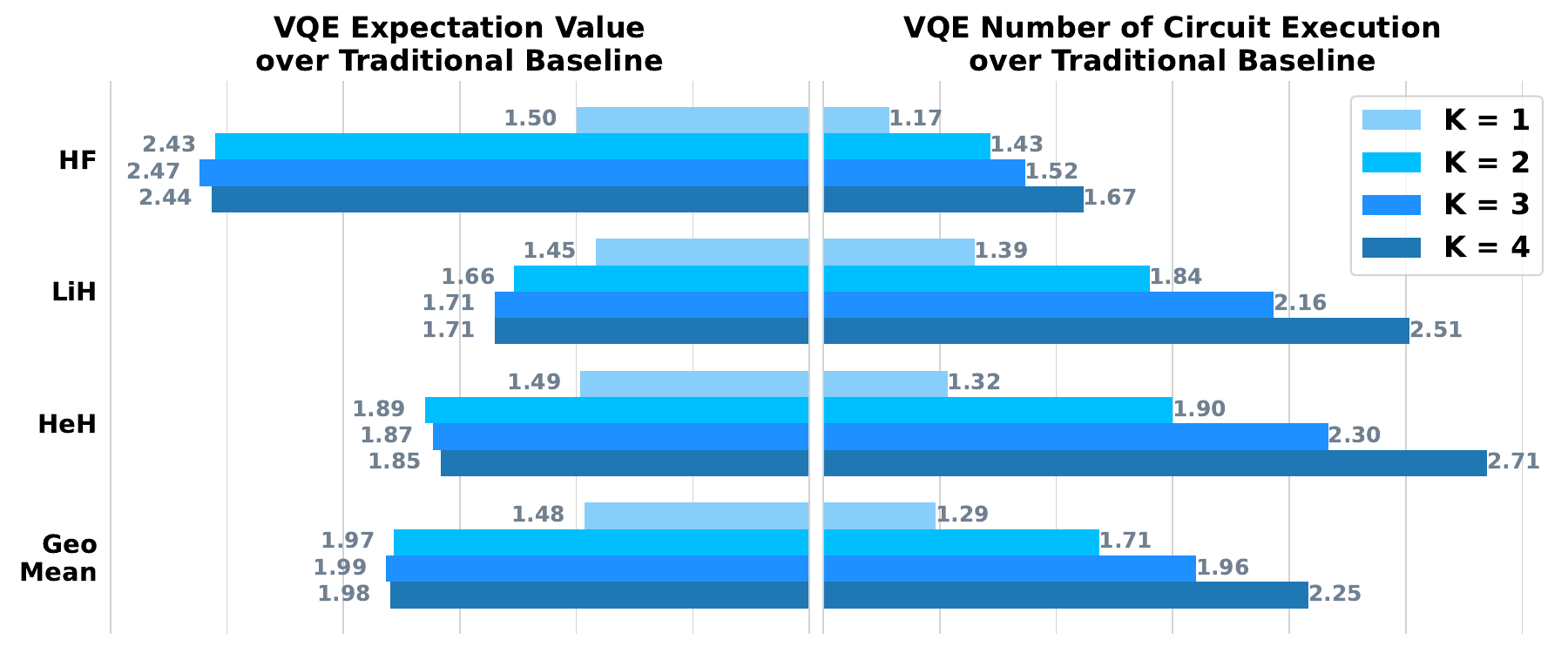}
  \vspace{-20pt}
  \caption{Simulation evaluation of \name using $K$ reference circuits.}
  % \vspace{-1.5em}
  \label{fig:ref}
\end{figure}
To further investigate the optimal $\text{TH}_p$ for molecules, four different thresholds (50, 70, 80, and 90) are analyzed along with Baseline in Fig.\ref{fig:threshold}. The evaluation is conducted on simulation across three molecules: HeH$^+$, LiH, and HF (Note: 60 has the same number of OC as 70 in HeH$^+$ and LiH molecules, while HeH$^+$ maintains a consistent OC count across these thresholds).
The 50 and 70 percent thresholds skip fewer impactful noise drift instances, resulting in relatively poor results. Although the 90 threshold yields similar improvements as 80, it requires more circuits to be executed. Consequently, the optimal threshold for selected molecules is found to be 80, which achieves strong benefits over Baseline while minimizing computing resources. Note that intelligent dynamic thresholding for different molecules may further improve benefits, but that is beyond the scope of this study.

\subsubsection{\textbf{Sweeping over \itname References Number}} \label{sec:eva_ref}

Section \ref{sec:reference} discusses using multi-references to improve the reliability of noise detection. Four different $K$ values (number of incorporated references) are analyzed alongside Baseline in Fig.\ref{fig:ref} for three molecules HeH$^+$, LiH, and HF.

For all the selected molecules, $K=1$ is a conservative scheme, which accepts some detrimental noise drift instances, and thus performs negative impacts on the VQA trace. 
For HeH$^+$ ion, incorporating three or four references for noise drift detection pushes \name to be worse than two references, since some iterations with tiny noise drift errors are avoided unnecessarily. Incorporating three references for molecules HF and LiH provides a good trade-off, achieving a 2.47$\times$ and 1.71$\times$ improvement over Baseline, respectively.

\section{RELATED WORK}
To contribute to the success of quantum computing during the NISQ era, it is critical to comprehend and control sources of noise, which typically include thermal fluctuations \cite{fluctuations} and magnetic flux \cite{Low_frequency, trapionchallenge, ZeemanIon}. Other sources, such as cosmic rays \cite{cosmic_rays}, device defects \cite{decoherence_dielectric_loss}, quantum drift \cite{quantum_drift} and external stimuli \cite{external1, external2, externalerror} also cause the pernicious effects on qubits. 

By understanding noise, multiple works from different hierarchies have investigated both error correction \cite{error_correction, error_correction_2, error_correction_3, error_correction_4} and error mitigation techniques to efficiently drive the computing power of quantum \cite{miti_3, miti_4, miti_5, miti_7, miti_8, miti_9}. Some works target specific noise sources and propose corresponding solutions, such as crosstalk-oriented compilation methods \cite{miti_2} and experiment-based techniques for measurement error \cite{miti_1}. Beyond these works, mapping virtual to physical qubits is also a primary technique used to alleviate noise effects \cite{mappingMiti, quantumnas}. The works in \cite{EQC, ensemble} even ensemble diverse qubit mappings or multi-device mappings. Other proposed mitigation techniques include learning-based methods \cite{learningMiti, learningMiti2, wang2022quest}, co-design of training, noise robustness \cite{QEC, quantumnat}, circuit structure analysis-based one \cite{10133711}, pulse-level optimizations \cite{liang2022variational} and Quantum neuron network compression \cite{hu2022quantum, hu2023battle}. However, the preceding works all primarily focus on static noise.

Regarding dynamic noise, \cite{qismet} profiles spike-like noise impacts and proposes a method with reference circuits. Comparatively, our overhead-aware scheduling method can provide significantly more accurate detection in facing various types of noise drift (not limited to transient spikes) while heuristically minimizing required computing resources. 

\section{CONCLUSION}
\name, as proposed in this paper, takes proactive steps to address the detrimental impact of noise drifts and to neutralize the required execution overhead. To achieve this, \name incorporates multi-reference circuits to faithfully detect noise drift errors in VQA iterations in order to maintain the fidelity of the iterations in drift-free scenarios. Furthermore, \name deploys Pauli-term subsetting to replace reference circuits with their prime subset circuits and partition the circuit execution to efficiently reduce the corresponding circuit execution overhead by detecting noise drift.

\section*{ACKNOWLEDGMENT}

The work was funded in part by National Science Foundation (NSF) CNS-2112562,
in part by ARO W911NF-19-2-0107,
in part by STAQ Project (PHY-1818914), 
in part by EPiQC --- an NSF Expeditions in computing (CCF-1832377), 
in part by NSF Quantum Leap Challenge Institute for Robust Quantum Simulation (OMA-2120757), 
in part by NSF CROSS --- Cross-layer Coordination and Optimization for Scalable and Sparse Tensor Networks (CCF-2217020); 
in part by EPiQC, an NSF Expedition in Computing, under award CCF-1730449; in part by NSF award 2110860; 
in part by the US Department of Energy Office of Advanced Scientific Computing Research, Accelerated Research for Quantum Computing Program; 
in part by the NSF Quantum Leap Challenge Institute for Hybrid Quantum Architectures and Networks (NSF Award 2016136) and in part based upon work supported by the U.S. Department of Energy, Office of Science, National Quantum Information Science Research Centers. FTC is Chief Scientist for Quantum Software at Infleqtion and an advisor to Quantum Circuits, Inc.
We thank MIT-IBM Watson AI Lab, Qualcomm Innovation Fellowship for supporting this research. We acknowledge the use
of IBM Quantum services for this work.

\clearpage

\bibliographystyle{IEEEtran}
\bibliography{reference}

\end{document}